\documentclass[english]{article}
\usepackage[T1]{fontenc}
\usepackage[latin9]{inputenc}
\usepackage{geometry}
\usepackage{color}
\usepackage{babel}
\usepackage{array}
\usepackage{float}
\usepackage{url}
\usepackage{multirow}
\usepackage{amsmath}
\usepackage{graphicx}
\usepackage{setspace}
\usepackage[unicode=true,pdfusetitle,
 bookmarks=true,bookmarksnumbered=false,bookmarksopen=false,
 breaklinks=false,pdfborder={0 0 0},pdfborderstyle={},backref=false,colorlinks=false]
 {hyperref}

\makeatletter

\providecommand{\tabularnewline}{\\}
\floatstyle{ruled}
\newfloat{algorithm}{tbp}{loa}
\providecommand{\algorithmname}{Algorithm}
\floatname{algorithm}{\protect\algorithmname}

\newenvironment{lyxlist}[1]
	{\begin{list}{}
		{\settowidth{\labelwidth}{#1}
		 \setlength{\leftmargin}{\labelwidth}
		 \addtolength{\leftmargin}{\labelsep}
		 }}
	{\end{list}}

\usepackage{algorithm,algpseudocode,animate}

\algnewcommand\algorithmicforeach{\textbf{for each}}
\algdef{S}[FOR]{ForEach}[1]{\algorithmicforeach\ #1\ \algorithmicdo}
\algdef{SE}[SUBALG]{Indent}{EndIndent}{}{\algorithmicend\ }%
\algtext*{Indent}
\algtext*{EndIndent}

\usepackage{xcolor}

\@ifundefined{showcaptionsetup}{}{%
 \PassOptionsToPackage{caption=false}{subfig}}
\usepackage{subfig}
\makeatother

\begin{document}
\title{Efficient smoothed particle radiation hydrodynamics II:\\
 Radiation hydrodynamics\vspace{0.1\textheight}
}
\author{Brody R. Bassett$^{a,\star}$, J. Michael Owen$^{a}$, Thomas A. Brunner$^{a}$\\
\\
{\normalsize{}\href{mailto:bassett4@llnl.gov}{bassett4@llnl.gov},
\href{mailto:owen8@llnl.gov}{owen8@llnl.gov}, \href{mailto:brunner6@llnl.gov}{brunner6@llnl.gov}}\\
{\normalsize{}}\\
{\normalsize{}$^{a}$Lawrence Livermore National Laboratory}\\
{\normalsize{}7000 East Avenue, Livermore, CA, 94550}\\
{\normalsize{}}\\
{\normalsize{}$^{\star}$Corresponding author}\vspace{0.1\textheight}
}
\date{$ $}
\maketitle
\begin{abstract}
The radiation hydrodynamics equations for smoothed particle hydrodynamics
are derived by operator splitting the radiation and hydrodynamics
terms, including necessary terms for material motion, and discretizing
each of the sets of equations separately in time and space. The implicit
radiative transfer discussed in the first paper of this series is
coupled to explicit smoothed particle hydrodynamics. The result is
a multi-material meshless radiation hydrodynamics code with arbitrary
opacities and equations of state that performs well for problems with
significant material motion. The code converges with second-order
accuracy in space and first-order accuracy in time to the semianalytic
solution for the Lowrie radiative shock problem and has competitive
performance compared to a mesh-based radiation hydrodynamics code
for a multi-material problem in two dimensions and an ablation problem
inspired by inertial confinement fusion in two and three dimensions.
\\
\end{abstract}

\pagebreak{}

\section{Introduction}

\textcolor{red}{}

Radiation hydrodynamics adds important physics to the standard hydrodynamics
equations, including energy transfer by radiative processes and radiation
effects that alter the hydrodynamics equations directly. Radiation
typically transfers energy at speeds that far exceed those of fluid
flow or thermal conduction, and can noticeably affect the hydrodynamic
system at temperatures of thousands of degrees Kelvin. Due to the
dependence of radiation energy on the fourth power of the temperature,
when the temperature increases a few orders of magnitude beyond that,
the radiation additionally imparts significant momentum and pressure
to the fluid \cite{castor2004radiation}. The coupled radiation hydrodynamic
equations are necessary in order to study a wide variety of problems
in astrophysics such as star formation, stellar structure and evolution,
accretion around stellar and collapsed objects (such as neutron stars,
white dwarves, and black holes), supernova, active galactic nuclei,
galaxy formation, etc. Radiation hydrodynamics is also important for
high-energy density experiments, such as inertial confinement fusion
research \cite{nuckolls1972laser,lindl1995development}. In this paper
radiation hydrodynamics is numerically modeled with the assumption
that the radiative and material energies are significantly (but non-trivially)
coupled, i.e. when simplifying assumptions such as the matter is either
decoupled from the radiation (free-streaming) or so tightly bound
that the photons are simply another component of the fluid and can
be accounted for by a modified equation of state are not appropriate.

Smoothed particle hydrodynamics (SPH) is a meshfree method typically
formulated for hydrodynamics only, i.e., without including the effects
of radiation. In the first paper of this series (hereafter Paper I)
an efficient flux-limited diffusion treatment for radiation (coupled
to the material energy) is derived based upon an SPH differencing
operator, which is therefore both meshfree and well-suited to combine
with SPH to create a meshfree radiation hydrodynamic method. Several
implementations of radiation within SPH are discussed in Paper I,
including ray tracing \cite{altay2008sphray} and Monte Carlo methods
\cite{nayakshin2009dynamic}. Flux-limited diffusion, the approximation
used here, remains the most common method due to its simplicity. Early
efforts using ideal gas laws and simple, iterative convergence schemes
\cite{whitehouse2004smoothed,whitehouse2005faster,mayer2007fragmentation},
which were later generalized to more difficult equations of state
and more robust solution methods \cite{viau2006implicit}, including
the similar optically-thin Variable Eddington Factor method \cite{petkova2009implementation}.
Many of the problems that have been tested in SPH flux-limited diffusion
are in 1D, and they do not always have a known solution. The 2D and
3D problems that have been studied are mostly \textcolor{black}{astrophysical problems such as star formation}
without a reference solution, although in at least one case a code-to-code
comparison is used to check the solution \cite{petkova2009implementation}. 

\textcolor{black}{The general equations of radiation hydrodynamics contain relativistic
terms due to the different reference frames of the fluid and radiation.
Approximations on the order of the fluid velocity over the speed of
light are generally applied to simplify the equations for problems
with nonrelativistic fluid velocities \cite{buchler1983radiation,castor2004radiation,morel2006discrete,mihalas2013foundations,lowrie2014simple}.
The radiation transport equation can then be integrated over angle
to produce the moment equations. The diffusion equations presented
in Ref. \cite{castor2004radiation} are those used in this paper.
Similar equations could be derived from the angularly-dependent transport
equation in Ref. \cite{morel2006discrete} or \cite{lowrie2014simple}.
While this paper assumes a simple relativistic approximations for
the radiation equations, there are more detailed treatments tailored
to diffusion that may provide additional accuracy in certain regimes
\cite{krumholz2007equations}. }Radiation hydrodynamics has long been studied in the context of meshed
methods (such as Eulerian, Lagrangian, AMR, and ALE methods; see the
extensive discussion in \cite{castor2004radiation}), but comparatively
little work has been done on meshfree methods such the one here. 

In Paper I, solution methods for the coupled radiation diffusion and
material energy equations for SPH are derived and verified. In Sec.
\ref{sec:theory} of this paper, the equations of radiation hydrodynamics
are introduced and applied to SPH, including a fast and robust time
discretization scheme that limits the number of iterations needed
to converge the system. Sec. \ref{sec:methodology} briefly discusses
the implementation of the methods described in Paper I and Sec. \ref{sec:methodology},
which allows for the simulation of large systems relatively quickly
on distributed architectures. In Sec. \ref{sec:results} multiple
tests of the coupled radiation hydrodynamics method are presented,
including a radiation hydrodynamics shock problem with a semi-analytic
solution for comparison, as well as several multidimensional examples
where the results are compared against a meshed method, including
a problem with multiple materials and an ablation problem inspired
by inertial confinement fusion.

\section{Theory\label{sec:theory}}

In this section, the radiation hydrodynamics equations are introduced
and discretized. The radiation hydrodynamics equations are kept general
to permit an arbitrary equation of state. The equations are operator
split to separate the thermal radiative transfer equations from the
hydrodynamics equations, which isolates the radiation solve. The hydrodynamics
equations are discretized explicitly in time, while the radiation
equations are discretized implicitly in time. Finally, SPH spatial
differencing is applied to the equations. 

\subsection{The radiation hydrodynamics equations\label{subsec:rad-hydro-equations}}

The radiation hydrodynamics equations are \begin{subequations}
\begin{gather}
\frac{D\rho}{Dt}=-\rho\partial_{x}^{\alpha}v^{\alpha},\label{eq:mass-conservation}\\
\rho\frac{Dv^{\alpha}}{Dt}=-\partial_{x}^{\alpha}p+\frac{\sigma_{t}}{c}F^{\alpha},\label{eq:momentum}\\
\rho\frac{De}{Dt}=-\partial_{x}^{\alpha}pv^{\alpha}-c\sigma_{a}B+c\sigma_{a}E+Q_{e},\\
\frac{DE}{Dt}=-E\partial_{x}^{\alpha}v^{\alpha}-P^{\alpha\beta}\partial_{x}^{\alpha}v^{\beta}-\partial_{x}^{\alpha}F^{\alpha}-c\sigma_{a}E+c\sigma_{a}B+Q_{E},\label{eq:zeroth-moment}\\
\frac{1}{c}\frac{DF^{\alpha}}{Dt}=-\frac{1}{c}F^{\alpha}\partial_{x}^{\beta}v^{\beta}-c\partial_{x}^{\beta}P^{\alpha\beta}-\sigma_{t}F^{\alpha},\label{eq:first-moment}
\end{gather}
\end{subequations}with the variables
\begin{lyxlist}{00.00.0000}
\item [{$t$,}] time,
\item [{$x$,}] position,
\item [{$v^{\alpha}$,}] velocity,
\item [{$\rho$,}] mass density,
\item [{$p$,}] pressure,
\item [{$e$}] specific material energy,
\begin{spacing}{0.7}
\item [{$E$,}] radiation energy density,
\end{spacing}
\item [{$F^{\alpha}$,}] radiation flux, 
\item [{$P^{\alpha\beta}$,}] radiation pressure,
\begin{spacing}{0.7}
\item [{$T$,}] material temperature,
\end{spacing}
\item [{$B$,}] integrated photon emission function,
\item [{$c$,}] speed of light in a vacuum,
\begin{spacing}{0.7}
\item [{$\sigma_{t}$,}] total opacity,
\item [{$\sigma_{a}$,}] absorption opacity,
\item [{$a$,}] black-body constant,
\item [{$Q_{e}$,}] nonhomogeneous material energy source,
\item [{$Q_{E}$,}] nonhomogeneous radiation energy source,
\end{spacing}
\end{lyxlist}
and the Lagrangian (or material) derivative,
\begin{equation}
\frac{D}{Dt}=\partial_{t}+v^{\alpha}\partial_{x}^{\alpha}.
\end{equation}
\textcolor{black}{As in Part I, Greek letters used as superscripts indicate dimensional
components of a vector and repeated indices indicate summation.}

The same energy and angular approximations apply as in Part I, namely
that the transport equation has been integrated over all energy frequencies
and over angle to produce the first two moments. The emission function
integrated over energy is 
\begin{equation}
B=aT^{4}.\label{eq:emission}
\end{equation}

To derive the diffusion equation, the time derivative terms in Eq.
(\ref{eq:first-moment}) are neglected, which results in Fick's Law,
\begin{equation}
F^{\alpha}=-\frac{c}{3\sigma_{t}}\partial_{x}^{\alpha}E.\label{eq:ficks}
\end{equation}
A flux limiter $\lambda$ is applied to prevent the radiation from
propagating faster than is physical (see Paper I for details). When
these two approximations are applied to Eq. (\ref{eq:zeroth-moment}),
the result in the flux-limited diffusion equation, 
\begin{equation}
\frac{DE}{Dt}=-\frac{4}{3}E\partial_{x}^{\alpha}v^{\alpha}+\partial_{x}^{\alpha}\left(\frac{c\lambda}{\sigma_{t}}\partial_{x}^{\alpha}E\right)-c\sigma_{a}E+c\sigma_{a}B+Q_{E}.\label{eq:diffusion}
\end{equation}
The same approximations are applied to the radiation momentum term
{[}Eq. (\ref{eq:momentum}){]},
\begin{equation}
\rho\frac{Dv^{\alpha}}{Dt}=-\partial_{x}^{\alpha}p-\lambda\partial_{x}^{\alpha}E.
\end{equation}
The flux limiter in this equation prevents the radiation pressure
from unphysically exceeding $E$.

\subsection{Time discretization}

Two separate operator split procedures are performed on the coupled
radiation hydrodynamics equations to make their solution more tractable.
These include splitting the work term from the diffusion equation
and splitting the hydrodynamics from the radiative transfer, which
also has the effect of splitting the Lagrangian time derivative for
the radiative transfer equations. The time discretization is done
separately for the hydrodynamics equations, the radiation work equation,
and the TRT equations. The hydrodynamics time update is performed
explicitly. The radiation work update is performed based on the explicit
hydrodynamics update. Finally, the thermal radiative transfer update
is performed implicitly. 

\subsubsection{Radiation work}

The zeroth-moment equation {[}Eq. (\ref{eq:zeroth-moment}){]} with
the Lagrangian derivative contains velocity-related terms that are
required to be accurate to order $v/c$ with material motion \cite{castor2004radiation},
\begin{equation}
\partial_{x}^{\alpha}v^{\alpha}E+P^{\alpha\beta}\partial_{x}^{\alpha}v^{\beta}.
\end{equation}
The term $v^{\alpha}\partial_{x}^{\alpha}E$, which is included in
the Lagrangian derivative, serves to entrain the radiation with the
fluid flow. The other terms, $E\partial_{x}^{\alpha}v^{\alpha}+P^{\alpha\beta}\partial_{x}^{\alpha}v^{\beta},$account
for the work done in compression or expansion of the radiation energy
density. For the diffusion equation, including the time derivative,
this term is simplified to 
\begin{equation}
\frac{DE}{Dt}+\frac{4}{3}E\partial_{x}^{\alpha}v^{\alpha}.
\end{equation}
The velocity term is operator split, which produces a standard Lagrangian
radiation diffusion equation,
\begin{equation}
\frac{DE}{Dt}=\partial_{x}^{\alpha}\left(\frac{c\lambda}{\sigma_{t}}\partial_{x}^{\alpha}E\right)-c\sigma_{a}E+c\sigma_{a}B+Q_{E},\label{eq:diffusion-presplit}
\end{equation}
along with an equation that calculates work done on the radiation
field,
\begin{equation}
\frac{DE}{Dt}=-\frac{4}{3}E\partial_{x}^{\alpha}v^{\alpha}.
\end{equation}
Using the mass conservation equation {[}Eq. (\ref{eq:mass-conservation}){]},
the work equation can be written as 
\begin{equation}
\frac{1}{E}\frac{DE}{Dt}=\frac{4}{3}\frac{1}{\rho}\frac{D\rho}{Dt}.\label{eq:work}
\end{equation}

The radiation work equation {[}Eq. (\ref{eq:work}){]} is discretized
in time by integrating from the start $t_{n-1}$ to the end $t_{n}$
of a time step,
\begin{equation}
\int_{t_{n-1}}^{t_{n}}\frac{1}{E}\frac{DE}{Dt}dt=\frac{4}{3}\int_{t_{n-1}}^{t_{n}}\frac{1}{\rho}\frac{D\rho}{Dt}dt,
\end{equation}
which can be performed analytically,
\begin{equation}
\ln E^{n}-\ln E^{n-1}=\frac{4}{3}\left(\ln\rho^{n}-\ln\rho^{n-1}\right),
\end{equation}
and simplified by exponentiation to
\begin{equation}
\frac{E^{n}}{E^{n-1}}=\left(\frac{\rho^{n}}{\rho^{n-1}}\right)^{4/3}.\label{eq:radiation-work}
\end{equation}
This is similar to the treatment of work in Ref. \cite{bowers1991numerical}.
After the hydrodynamics update has been performed, the ratio of the
mass densities is used to update the radiation energy density before
the thermal radiative transfer update.

\subsubsection{Radiation and hydrodynamics operator split}

SPH hydrodynamics is traditionally solved using explicit time steps,
but explicit time steps for radiation diffusion would be cost prohibitive
due to the small time step needed to resolve the speed of propagation
for radiation using the Courant-Friedrichs-Lewy (CFL) condition, which
limits the propagation of a signal within each time step to some fraction
of the element size. To solve the hydrodynamics equations explicitly
and the diffusion equations implicitly, the two must be operator split. 

This second operator split produces two sets of equations, the hydrodynamics
equations (including a term for the radiative momentum deposition)
which are advanced explicitly in time, \begin{subequations}\label{eq:hydro-split}
\begin{gather}
\frac{D\rho}{Dt}=-\rho\partial_{x}^{\alpha}v^{\alpha},\\
\rho\frac{Dv^{\alpha}}{Dt}=-\partial_{x}^{\alpha}p-\lambda\partial_{x}^{\alpha}E,\label{eq:momentum-split}\\
\rho\frac{De}{Dt}=-\partial_{x}^{\alpha}pv^{\alpha},\\
\frac{DE}{Dt}=0,\label{eq:radiation-entrainment}
\end{gather}
\end{subequations}and the coupled material and radiation energy equations
which are advanced implicitly in time,\begin{subequations}\label{eq:rad-split}
\begin{gather}
\rho\partial_{t}e=-c\sigma_{a}B+c\sigma_{a}E+Q_{e},\\
\partial_{t}E=\partial_{x}^{\alpha}\left(\frac{c\lambda}{\sigma_{t}}\partial_{x}^{\alpha}E\right)-c\sigma_{a}E+c\sigma_{a}B+Q_{E}.
\end{gather}
\end{subequations}The hydrodynamics equations remain in the material
frame of reference with Lagrangian derivatives, as needed for SPH.
The thermal radiative transfer (TRT) equations are in the lab frame
of reference with Eulerian derivatives, which allows for their solution
without consideration of material motion. As described in Sec. \ref{subsec:hydrodynamics-time},
the inclusion of Eq. (\ref{eq:radiation-entrainment}) is a formalism
and does not require implementation. 

\subsubsection{Hydrodynamics\label{subsec:hydrodynamics-time}}

The hydrodynamics equations are discretized explicitly in time. Equation
(\ref{eq:radiation-entrainment}), which states that the radiation
energy moves along with the fluid, comes for free in a Lagrangian
method if the radiation and hydrodynamics discretizations use the
same points, as is the case here. During the explicit hydrodynamics
update, the radiation energy density is moved along with the fluid,
which satisfies Eq. (\ref{eq:radiation-entrainment}). This leaves
a standard set of hydrodynamics equations, which is discretized using
standard SPH methodology. 

Each of the hydrodynamics equations {[}Eqs. (\ref{eq:hydro-split}){]}
(except for the radiation energy density equation) is written in the
form 
\begin{equation}
\frac{Du}{Dt}=g\left(t,u\right),
\end{equation}
and discretized explicitly. For example, second-order Runge-Kutta,
which is used for the results in this paper, has the form\begin{subequations}
\begin{gather}
k_{1}=\Delta tg\left(t_{n},u^{n}\right),\\
k_{2}=\Delta tg\left(t_{n}+\frac{\Delta t}{2}u^{n}+\frac{1}{2}k_{1}\right),\\
u^{n+1}=u^{n}+k_{2}.
\end{gather}
\end{subequations}The radiation momentum term $-\lambda\partial_{x}^{\alpha}E$
{[}Eq. (\ref{eq:momentum-split}){]} is lagged to always use the value
$E^{n-1}$. 

\subsubsection{Thermal radiative transfer}

The time discretization of the thermal radiative transfer equations
is discussed at length in Paper I. In summary, Eqs. (\ref{eq:rad-split})
are discretized in time using backward Euler, and solved using Newton
iteration. A Schur complement is used to eliminate part of the Jacobian,
which results in the radiation diffusion equation\begin{subequations}\label{eq:time-discretized-trt}
\begin{equation}
\dfrac{1}{\Delta t}E^{\ell+1}-\partial_{x}^{\alpha}\left(\dfrac{c\lambda}{\sigma_{t}}\partial_{x}^{\alpha}E^{\ell+1}\right)+c\sigma_{a}fE^{\ell+1}=\dfrac{1}{\Delta t}E^{n-1}+c\sigma_{a}B-\left(1-f\right)c\sigma_{a}E^{\ell}+Q_{E}^{n},\label{eq:time-discretized-diffusion}
\end{equation}
where $\ell$ is the iteration index, $f$ is the Fleck factor \cite{fleckjr1971implicit},
\begin{equation}
f=\left(1+c\sigma_{a}\Delta t\frac{4aT^{3}}{\rho c_{v}}\right)^{-1},
\end{equation}
and $c_{v}$ is the specific heat. The $E$ values are all at the
current time step $n$ except for $E^{n-1}$. The solution of Eq.
(\ref{eq:time-discretized-diffusion}) represents a single inexact
Newton iteration. The material energy equation,
\begin{equation}
\dfrac{\rho}{\Delta t}\left(e-e^{n-1}\right)+c\sigma_{a}B=c\sigma_{a}E+Q_{e}^{n},
\end{equation}
\end{subequations}is solved separately to convergence using a separate
Newton iteration process before each diffusion solve. The solution
of these equations (described in Paper I, Alg. 1) is performed after
the hydrodynamics and radiation work equations have been updated.
The opacities and other material parameters ($\sigma_{a}$, $f$,
$D$, $\lambda$, $c_{v}$, and $\rho$) are calculated based on the
values from the end of the hydrodynamics update. The flux limiter,
which depends on $E$, is updated after the radiation work update
of $E$. 

\subsubsection{Coupled system}

The time integration procedure for the coupled radiation hydrodynamics
equations is presented in Alg. \ref{alg:rad-hydro-time}. For the
hydrodynamics update, the radiation momentum term is held constant
at the start-of-step value, $E^{n-1}$. Once the hydrodynamics update
is complete, the radiation work is calculated. This updated radiation
energy is then used in the thermal radiative transfer update. Before
the radiation update, the opacities, flux limiter, specific heat,
the Fleck factor are evaluated again using the updated hydrodynamics
variables and the radiation energy after the work equation is applied.
They are then held constant over the update, which means that the
preconditioner for the diffusion solver only needs to be initialized
once. 

The material energy is updated twice, first during the hydrodynamics
update and again during the radiation update. During the radiation
update, the material energy equation {[}Eq. (\ref{eq:discretized-materiale}){]}
should include the energy gained or lost over the hydrodynamics update.
This is done by providing this energy as a source to the material
energy equation,
\begin{equation}
Q_{e}^{n}=\frac{\rho}{\Delta t}\Delta e_{\text{hydro}}^{n}=\frac{\rho}{\Delta t}\left(e_{\text{hydro}}^{n}-e^{n-1}\right).\label{eq:material-energy-change}
\end{equation}

\subsection{Spatial discretization}

The spatial discretization of the radiation hydrodynamics equations
is performed using SPH, as described in Paper I. In review, the equations
are interpolated using a kernel, $W\left(x-x^{\prime},h\right)$,
with the notation
\begin{align}
\left\langle g\left(x\right)\right\rangle  & =\int_{V}W\left(x-x^{\prime},h\right)g\left(x\right)dV'\nonumber \\
 & \approx\sum_{j}V_{j}W\left(x-x_{j},h\right)g\left(x_{j}\right).
\end{align}
Derivatives are approximated similarly,
\begin{align}
\left\langle \partial_{x}^{\alpha}g\left(x\right)\right\rangle  & =\int_{V}W\left(x-x^{\prime},h\right)\partial_{x^{\prime}}^{\alpha}g\left(x^{\prime}\right)dV'\nonumber \\
 & =-\int_{V}\partial_{x^{\prime}}^{\alpha}W\left(x-x^{\prime},h\right)g\left(x^{\prime}\right)dV'\nonumber \\
 & =\partial_{x}^{\alpha}\int_{V}W\left(x-x^{\prime},h\right)g\left(x^{\prime}\right)dV'-g\left(x\right)\partial_{x}^{\alpha}\int_{V}W\left(x-x^{\prime},h\right)dV'\nonumber \\
 & \approx\sum_{j}V_{j}\left(g_{j}-g\left(x\right)\right)\partial_{x}^{\alpha}W\left(x-x_{j},h\right).\label{eq:sph-derivative}
\end{align}
Note that the $g\left(x\right)$ term is zero in the continuous derivative
(since $W$ is an even function), but is not necessarily zero in the
discrete approximation, as the quadrature points $x_{j}$ are not
always symmetric. This term ensures that the discrete derivative is
zero for a constant function. 

The equations without spatial derivatives are multiplied by $W\left(x_{i}-x^{\prime},h\right)$,
integrated, and then evaluated using the approximation $\left\langle g\left(x\right)\right\rangle _{i}\approx g_{i}$.
For the radiation work equation, this means the interpolated value,
\begin{equation}
\left\langle \frac{E^{n}}{E^{n-1}}\right\rangle _{i}=\left\langle \left(\frac{\rho^{n}}{\rho^{n-1}}\right)^{4/3}\right\rangle _{i},
\end{equation}
becomes a simple evaluation at point $i$, 
\begin{equation}
\frac{E_{i}^{n}}{E_{i}^{n-1}}=\left(\frac{\rho_{i}^{n}}{\rho_{i}^{n-1}}\right)^{4/3},
\end{equation}
which means the update in time can be done independently for each
point. 

\subsubsection{Hydrodynamics}

The spatial discretization of the hydrodynamics equations {[}Eqs.
(\ref{eq:hydro-split}){]} is similar, but includes derivatives in
addition to the values evaluated at specific points. To derive the
SPH form of the mass conservation equation, the original equation
is multiplied by the kernel centered at point $x_{i}$, $W\left(x_{i}-x^{\prime},h\right)$,
and integrated,
\begin{equation}
\left\langle \frac{1}{\rho\left(x^{\prime}\right)}\frac{D\rho\left(x^{\prime}\right)}{Dt}\right\rangle _{i}=-\left\langle \partial_{x}^{\alpha}v^{\alpha}\left(x^{\prime}\right)\right\rangle _{i}.
\end{equation}
Using the approximation $\left\langle g\left(x\right)\right\rangle _{i}\approx g_{i}$
for the time derivative term and the derivative approximation in Eq.
(\ref{eq:sph-derivative}) for the velocity derivative, the equation
becomes
\begin{equation}
\frac{1}{\rho_{i}}\frac{D\rho_{i}}{Dt}=-\sum_{j}V_{j}\left(v_{j}^{\alpha}-v_{i}^{\alpha}\right)\partial_{x_{i}}^{\alpha}W\left(x_{i}-x_{j},h\right)dV'.
\end{equation}
Performing similar manipulations to the other hydrodynamics equations
results in a standard set of SPH equations \cite{monaghan2005smoothed},
\begin{subequations}
\begin{gather}
\frac{D\rho_{i}}{Dt}=\rho_{i}\sum_{j}V_{j}\left(v_{i}^{\alpha}-v_{j}^{\alpha}\right)\partial_{x_{i}}^{\alpha}W_{ij},\label{eq:continuity}\\
\frac{Dv_{i}^{\alpha}}{Dt}=-\sum_{j}m_{j}\left(\frac{p_{j}}{\rho_{j}^{2}}+\frac{p_{i}}{\rho_{i}^{2}}\right)\partial_{x_{i}}^{\alpha}W_{ij},\\
\frac{De_{i}}{Dt}=-\sum_{j}m_{j}\left(\frac{p_{i}v_{j}^{\alpha}}{\rho_{i}^{2}}+\frac{p_{j}v_{i}^{\alpha}}{\rho_{j}^{2}}\right)\partial_{x_{i}}^{\alpha}W_{ij},\label{eq:emathydro}
\end{gather}
\end{subequations}where 
\[
W_{ij}=W\left(x_{i}-x_{j},h\right).
\]

Note that for the examples shown in this paper, the mass density $\rho_{i}$
is advanced using the SPH summation form,
\begin{equation}
\rho_{i}=\sum_{j}m_{j}W_{ij},\label{eq:sumdensity}
\end{equation}

rather than the continuity relation of Eq. (\ref{eq:continuity}),
which is only used to form the half-step estimate of the mass density
during the Runge-Kutta time advancement of the mass density. Similarly
the material energy relation in Eq. (\ref{eq:emathydro}) is only
used for the half-step estimate, while the actual hydrodynamic work
is evaluated using the compatible energy update described in Ref.
\cite{owen2014compatibly}.

\subsubsection{Thermal radiative transfer}

The spatial discretization of the radiation equations {[}Eqs. \ref{eq:time-discretized-trt}{]}
is described in Paper I. The result of this discretization is the
equations \begin{subequations}
\begin{equation}
\frac{\rho_{i}}{\Delta t}\left(e_{i}^{n,\ell+1}-e_{i}^{n-1}\right)+c\sigma_{a,i}B_{i}^{n,\ell+1}-c\sigma_{a,i}E_{i}^{n,\ell+1}-Q_{e,i}^{n}=0,\label{eq:discretized-materiale}
\end{equation}
\begin{align}
 & \frac{1}{\Delta t}E_{i}^{n,\ell+1}-\sum_{j}V_{j}\left(D_{i}+D_{j}\right)\left(E_{i}^{n,\ell+1}-E_{j}^{n,\ell+1}\right)\frac{x_{ij}^{\alpha}}{x_{ij}^{\beta}x_{ij}^{\beta}}\partial_{x_{i}}^{\alpha}W_{ij}+c\sigma_{a,i}f_{i}E_{i}^{n,\ell+1}\nonumber \\
 & \qquad=\frac{1}{\Delta t}E_{i}^{n-1}+c\sigma_{a,i}B_{i}^{n,\ell}-\left(1-f_{i}\right)c\sigma_{a,i}E_{i}^{n,\ell}+Q_{E,i}^{n},
\end{align}
with the diffusion coefficient $D$ and Fleck factor $f$ evaluated
at point $i$, 
\begin{gather}
D_{i}=\frac{c\lambda_{i}}{\sigma_{t,i}},\\
f_{i}=\left(1+c\sigma_{a,i}\Delta t\frac{4aT_{i}^{3}}{\rho_{i}c_{v,i}}\right)^{-1}.
\end{gather}
\end{subequations}The diffusion equation is solved using GMRES with
the Hypre BoomerAMG preconditioner \cite{falgout2002hypre}. 

\section{Methodology\label{sec:methodology}}

The radiation hydrodynamics methods from this paper are implemented
in the SPH code Spheral, which is available at \url{https://github.com/jmikeowen/spheral}.
Note, however, that the publicly available version of Spheral does
not include the radiation evolution described here and in Paper I.
Please contact the authors for access to the thermal radiative transfer
code and the input scripts for the problems in Sec. \ref{sec:results}. 

As in Paper I, the Wendland C4 kernel is used for the results in this
paper,
\begin{equation}
W_{\text{wendland}}\left(x_{ij},h\right)=\begin{cases}
k\left(1-r\right)^{5}\left(5r+1\right), & r\equiv x_{ij}/h\leq1,\\
0, & \text{otherwise},
\end{cases}\label{eq:wendlandc4}
\end{equation}
where $k$ is a normalization constant. The standard SPH evaluation
of the kernel $\psi\left(r\right)$ uses a scalar $h$, 
\begin{equation}
r=\frac{\sqrt{\left(x^{\alpha}-x^{\alpha,\prime}\right)\left(x^{\alpha}-x^{\alpha,\prime}\right)}}{h}.
\end{equation}

Although Spheral incorporates ASPH (a tensor generalization of the
scalar smoothing scale $h$, as described in \cite{owen1998adaptive}),
the examples in this paper simply use the scalar smoothing scale of
SPH. The smoothing scale $h_{i}$ is spatially (point-wise) varying,
and is adapted such that each point samples approximately a constant
number of neighbors within the sampling kernel {[}Eq. (\ref{eq:wendlandc4}){]}.
The algorithm for adjusting the smoothing scale per point is described
in Ref \cite{owen2010asph}. \textcolor{black}{Unlike in Part I, this smoothing scale is recalculated at every time
step due to the movement of the SPH particles.}

The code is also run with the hydrodynamics in compatible energy mode,
as described in Ref. \cite{owen2014compatibly}, which makes the hydrodynamics
step exactly energy conserving. The artificial viscosity used for
the problems in this paper is based on the Monaghan and Gingold viscosity
\cite{monaghan1983shock}. 

\section{Results\label{sec:results}}

Three problems are considered, including a radiating shock with a
semianalytic solution, a multi-material problem with strong radiation
coupling, and an ablation problem. For the latter two problems, the
results are compared to Kull, a mesh-based, arbitrary Lagrangian-Eulerian
(ALE) inertial confinement fusion code at Lawrence Livermore National
Laboratory \cite{rathkopf2000kull}. All three problems use an ideal
gas equation of state, with \begin{subequations}
\begin{gather}
p=\left(\gamma-1\right)\rho e,\\
T=\frac{\left(\gamma-1\right)\mu m_{p}}{k_{B}}e,
\end{gather}
\end{subequations}where $\gamma$ is the heat capacity ratio, $\mu$
is the molecular weight, $m_{p}$ is the proton mass, and $k_{B}$
is the Boltzmann constant. 

\subsection{Lowrie radiating shock}

To test the coupled radiation hydrodynamics, the planar shock wave
from Lowrie and Edwards \cite{lowrie2008radiative} is considered.
The spatial profile of the solution is semianalytic and moves at a
constant velocity. This problem simply translates the semianalytic
solution with time across the simulated volume, so the initial conditions
can be obtained from the solutions described in \cite{lowrie2008radiative}.
Two cases are considered, a Mach 2 shock with constant opacities and
a Mach 45 shock with opacities that depend on density and temperature.
The parameters for these shocks are listed in Table (\ref{tab:lowrie-parameters}).
The equation of state is for an ideal gas, with the atomic mass calculated
from the heat capacity $c_{v}$. A radiation wave ahead of the shock,
called a precursor, heats the material up to pre-shock temperatures.
When the shock hits, the material is discontinuously heated up to
a peak temperature, known as the Zel'dovich spike \cite{zeldovich2012physics}.
Finally, the material cools through radiative losses to post-shock
temperatures. The Mach 2 case is a subcritical shock, meaning the
pre-shock temperature is lower than the post-shock temperature, whereas
the Mach 45 case is supercritical, meaning the pre-shock and post-shock
temperatures are equal.

Ten points at the left and the right side of the problem are chosen
to be boundary points with constant values of velocity, density, material
energy, and radiation energy. The positions are set to move with the
constant velocities assigned to the points. The smoothing parameter
is fixed to its initial value for these boundary points. As the support
of the kernel is less than ten points across, this mimics a constant
boundary condition for the remainder of the points not on the boundary. 

\begin{figure}
\begin{centering}
\subfloat[Mach 2]{\begin{centering}
\includegraphics[width=0.75\textwidth]{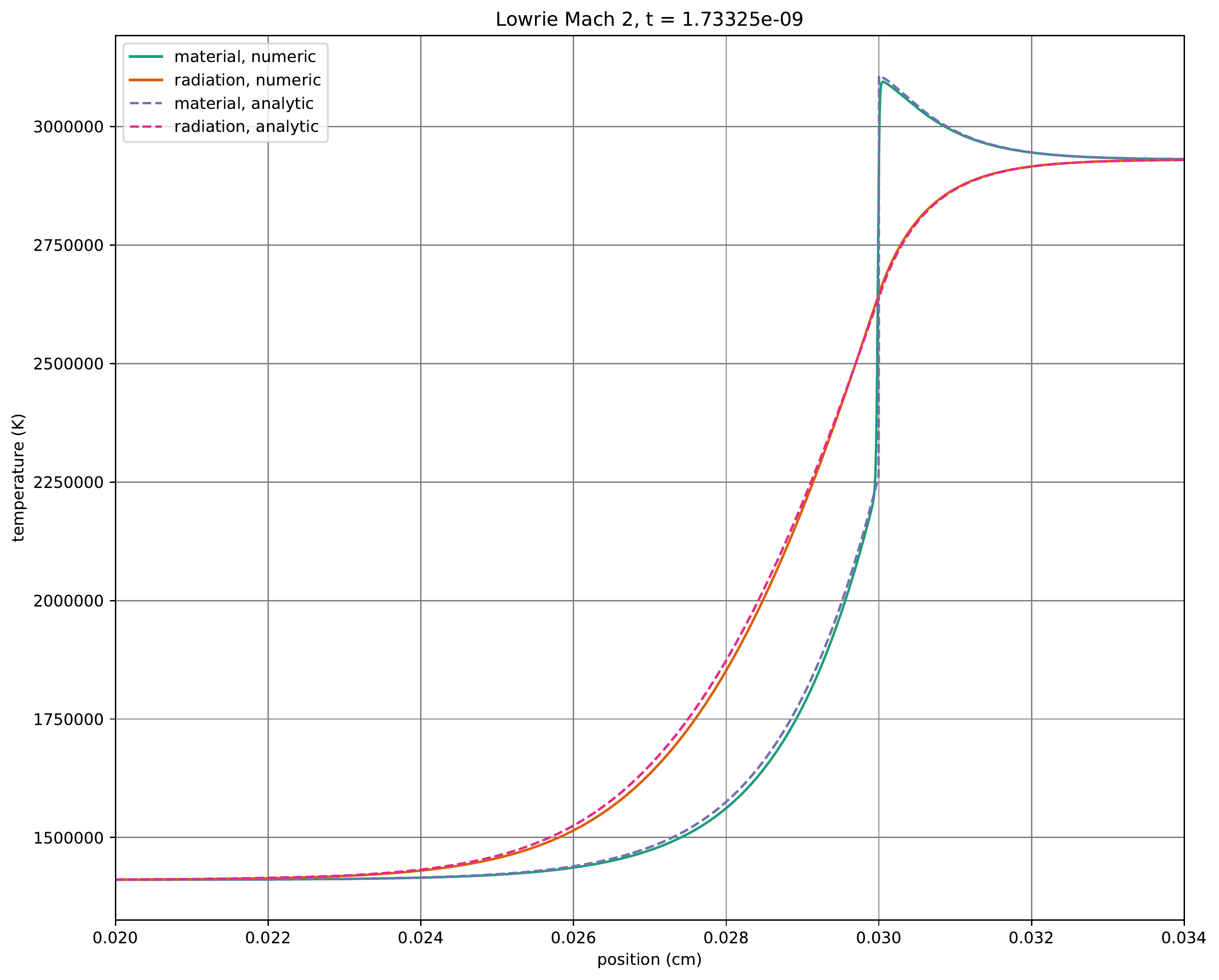}
\par\end{centering}
\label{fig:lowrie-2-solution}}
\par\end{centering}
\begin{centering}
\subfloat[Mach 45]{\begin{centering}
\includegraphics[width=0.75\textwidth]{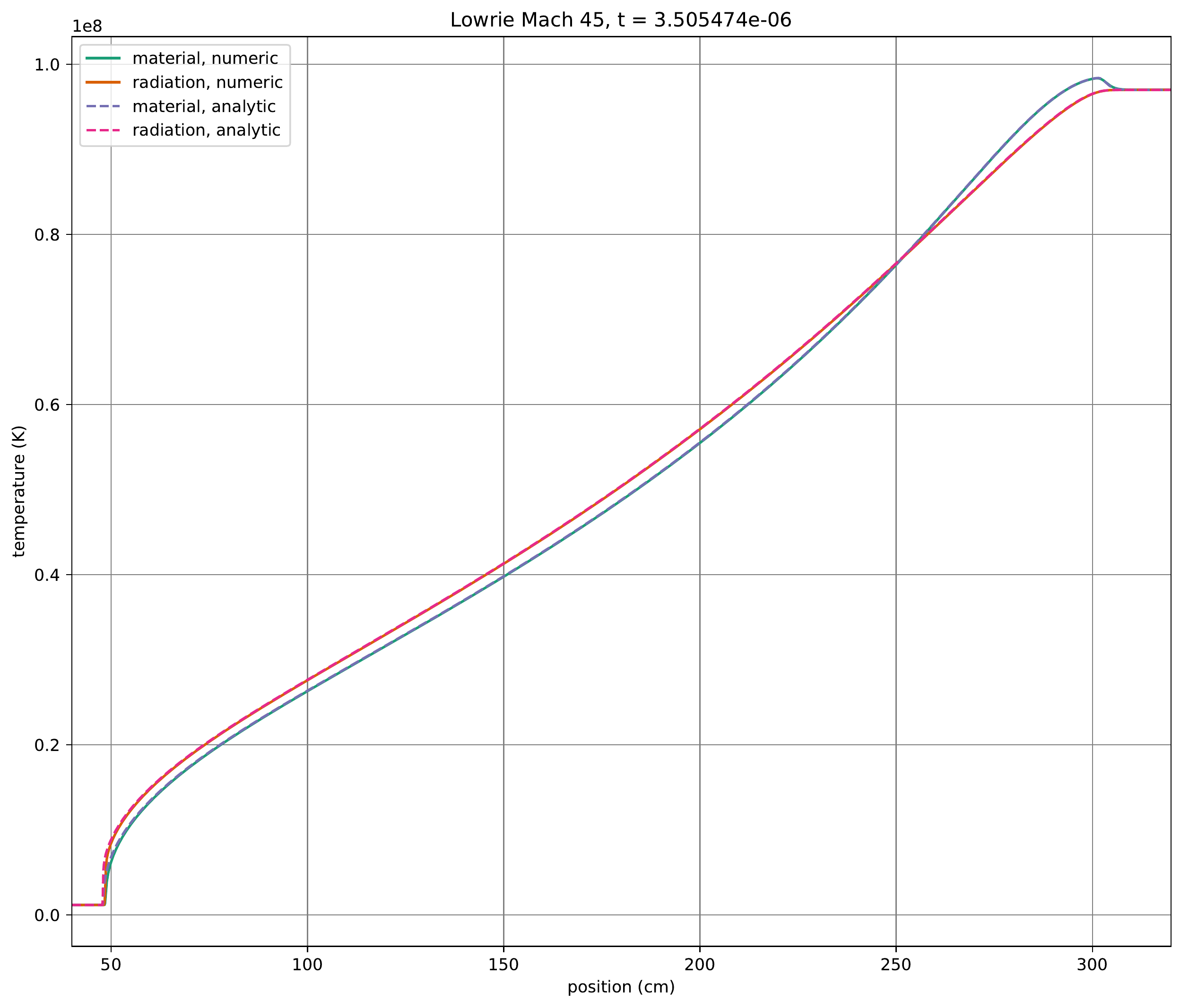}
\par\end{centering}
\label{fig:lowrie-45-solution}}
\par\end{centering}
\caption{Comparison of numerical and semianalytic material and radiation temperatures
for the Lowrie shock, 16,384 points.}
\end{figure}

\begin{figure}
\begin{centering}
\includegraphics[width=0.75\textwidth]{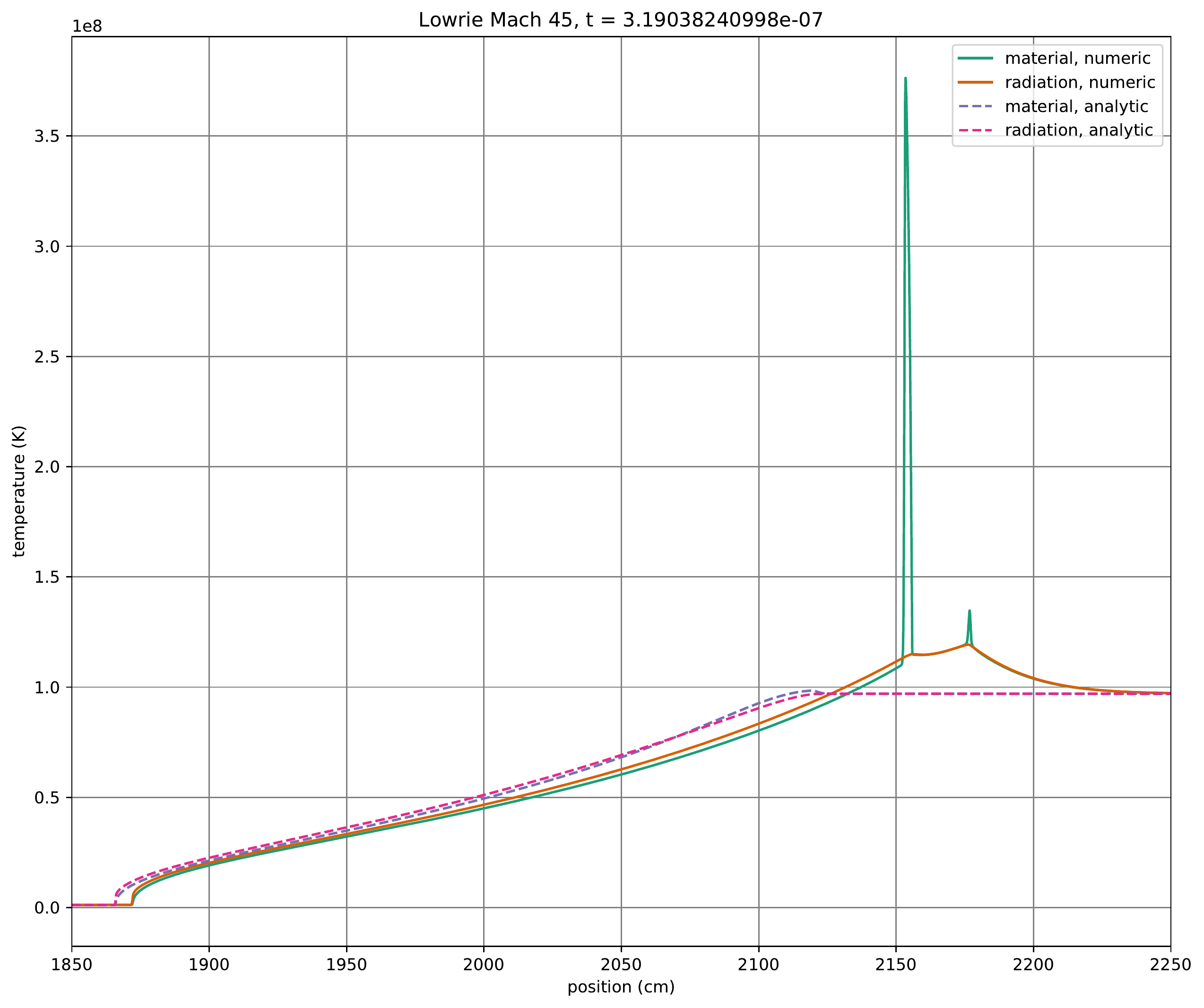}
\par\end{centering}
\caption{Comparison of numerical and semianalytic material and radiation temperatures
for the Lowrie shock without radiation momentum, 16,384 points. }
\label{fig:lowrie-45-solution-nomom}
\end{figure}

The solution for the Mach 2 case at the end time (after the shock
has traveled 0.06 cm) with 16384 total points is shown in Fig. \ref{fig:lowrie-2-solution}.
The radiation temperature is calculated based on Eq. (\ref{eq:emission}),
\begin{equation}
T=\left(\frac{E}{a}\right)^{1/4}.\label{eq:rad-temperature}
\end{equation}
The Zel'dovich spike is well-resolved, and the pre-shock and post-shock
temperatures agree with the semianalytic solution. The error appears
to be largest in the precursor region (around 0.2 percent $L_{\text{inf}}$
relative error), where the numerical solution appears to be more diffusive
than the semianalytic solution. For the Mach 45 case, also with 16384
points, the solution after the shock has traveled 2000 cm is shown
in Fig. \ref{fig:lowrie-45-solution}. The maximum relative error
in this case is 0.07 percent for the material and 0.01 percent for
the radiation, both of which occur near the shock front where the
temperature is smallest. For the Mach 45 case, the radiation momentum
term (see Sec. \ref{subsec:rad-hydro-equations}) is vital to calculating
the solution correctly. Without radiation momentum (which has the
same form as a hydrodynamic pressure), the shock lags behind the semianalytic
solution, causing a density spike that changes the spatial profile
of the solution (Fig. \ref{fig:lowrie-45-solution-nomom}). 

\begin{figure}
\begin{centering}
\subfloat[Mach 2]{\begin{centering}
\includegraphics[width=0.5\textwidth]{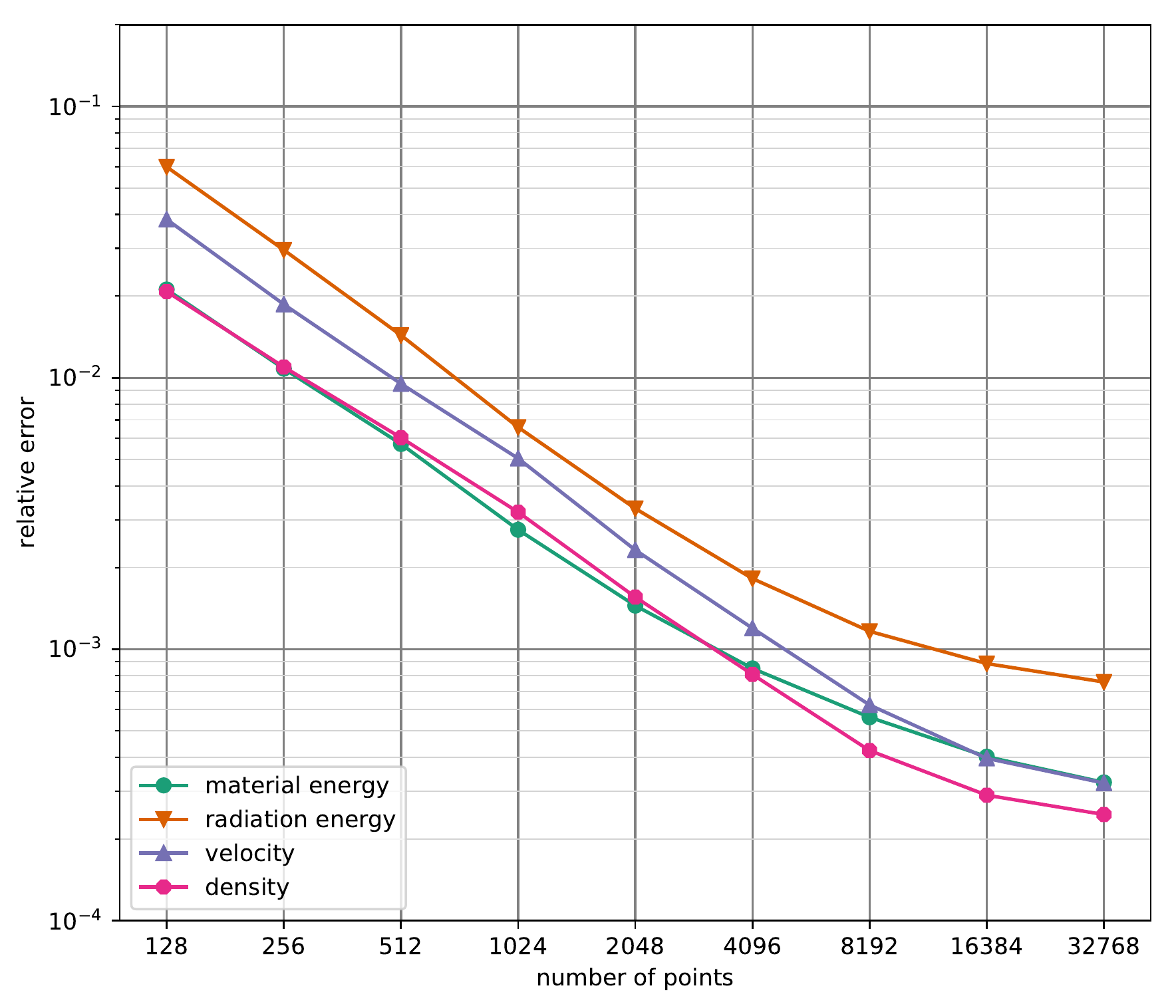}
\par\end{centering}
\label{fig:lowrie-convergence-2}}
\par\end{centering}
\begin{centering}
\subfloat[Mach 45]{\begin{centering}
\includegraphics[width=0.5\textwidth]{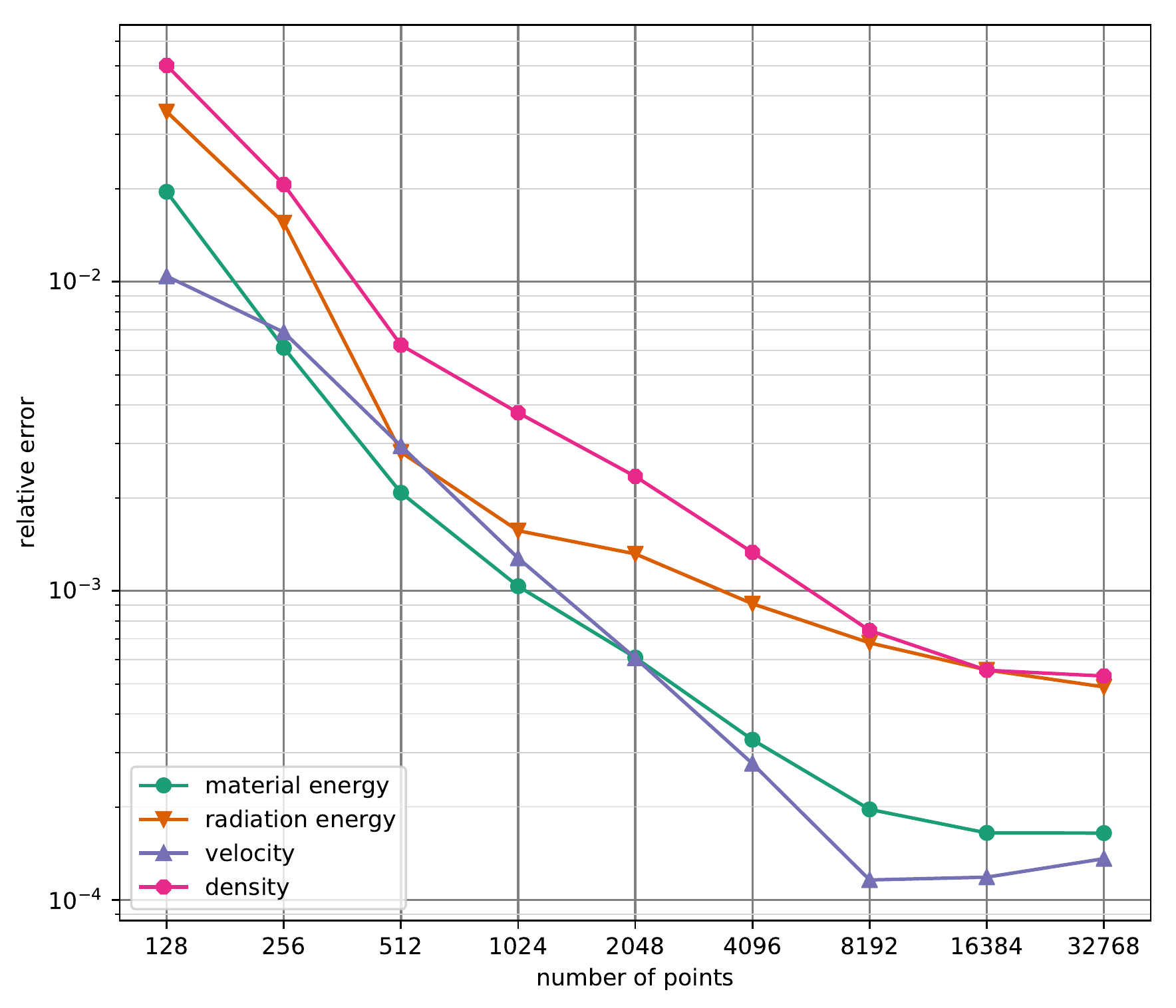}
\par\end{centering}
\label{fig:lowrie-convergence-45}}
\par\end{centering}
\caption{Convergence of the numerical result for the Lowrie shock to the semianalytic
solution.}
\end{figure}

Convergence results for the Mach 2 and Mach 45 cases are shown in
Figs. \ref{fig:lowrie-convergence-2} and \ref{fig:lowrie-convergence-45},
respectively. The numeric solution converges with second-order spatial
accuracy to the semianalytic solution. For a high number of points,
the error levels off due to the first-order radiation time discretization,
which is chosen for stability and not accuracy. The time step for
both Mach 2 and Mach 45 is limited by the CFL condition (see Sec.
\ref{subsec:hydrodynamics-time}), which decreases linearly as the
spacing between solution points is decreased, $\Delta t\propto\Delta x$.
This couples the temporal and spatial resolution, leading to an effective
first-order convergence unless the time step is manually set to lower
values. To see full second-order convergence in space, the time step
needs to decrease as $\Delta t\propto\Delta x^{2}$, but this becomes
cost-prohibitive for large simulations. For lower numbers of points,
the solutions converge with second-order accuracy despite the shock,
which is because the error is largest in the precursor region. 

\subsection{Triple point with radiation\label{subsec:triple-point}}

\begin{figure}
\begin{centering}
\includegraphics[width=1\textwidth]{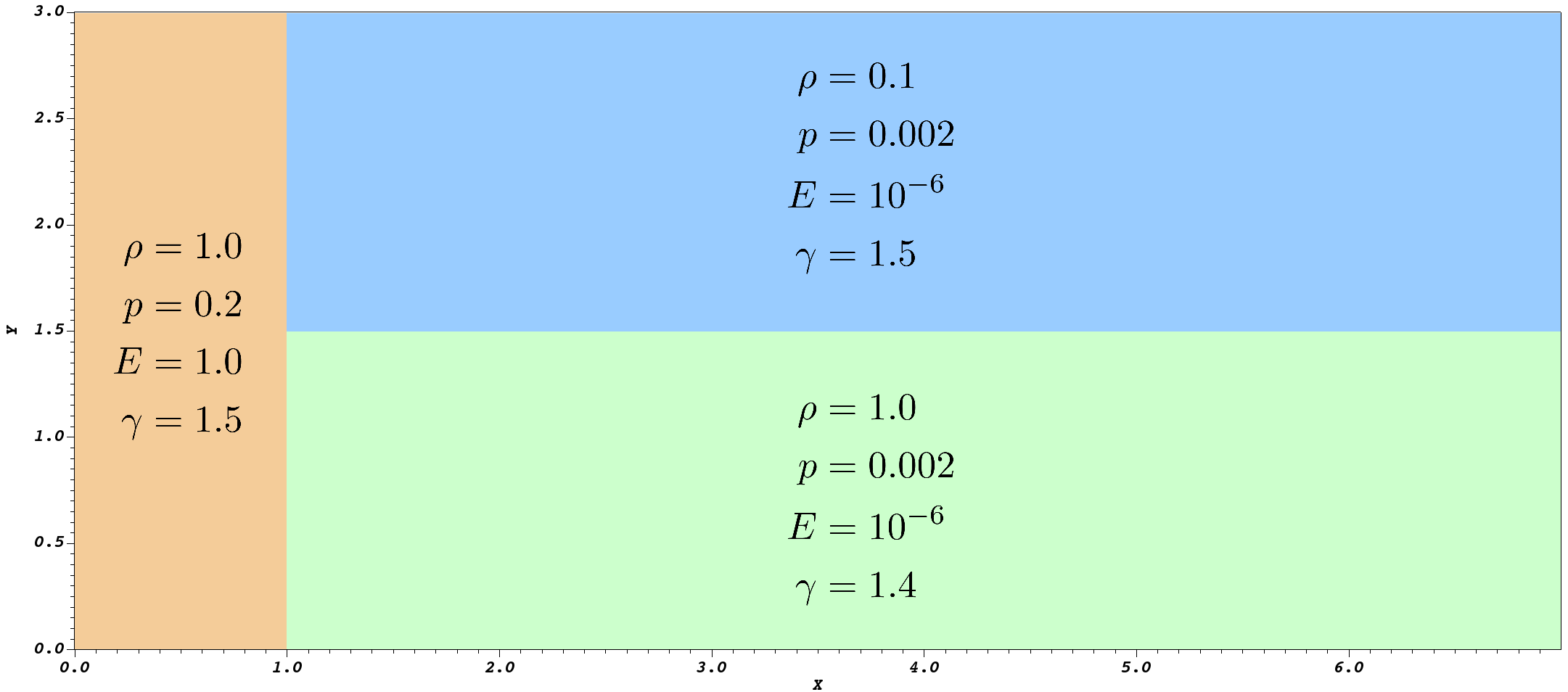}
\par\end{centering}
\caption{Initial conditions for the triple point problem, including density
$\rho$, pressure $p$, radiation energy density $E$, and the ratio
of specific heats, $\gamma$. The units are listed in Sec. \ref{subsec:triple-point}.}
\label{fig:triple-geom}
\end{figure}

The triple point problem is traditionally a hydrodynamics-only test
problem, and has been researched previously for various meshed methods
such as finite element methods \cite{dobrev2013high}, as well as
the meshfree conservative reproducing kernel (CRKSPH) example in \cite{frontiere2017crksph}.
The problem is changed here to start with significant radiation energy
in the high-pressure side, as shown in the initial conditions in Fig.
\ref{fig:triple-geom}. The pressure of the two low-pressure regions
is also decreased to allow the hydrodynamics to act on similar time
scales to the radiation hydrodynamical effects. Reflecting boundary
conditions are applied to each of the four boundary surfaces. The
unit system for Spheral is set to agree with the fixed unit system
of Kull, which is in terms of centimeters for length, grams for mass,
shakes for time (1 sh = $10^{-8}$ s), jerks for energy (1 jrk = $10^{16}$
erg), and keV for temperature (1 keV = $1.16045\times10^{7}$ K).
The initial conditions are also in this unit system (Fig. \ref{fig:triple-geom}),
with opacities of $\sigma_{a}=\sigma_{s}=100.0\rho\text{ cm}^{-1}$
in all regions of the problem. Similarly, while the ratio of specific
heats $(\gamma)$ is set per region as shown in Fig. \ref{fig:triple-geom},
the molecular weight is set to a constant value of $\mu=1$ in all
three regions. 

The problem is run in Spheral and Kull in low and high resolution,
with the run parameters in Table \ref{tab:triple-comparison}. In
Spheral, the points are mass-matched, meaning that in the low-density
region, the points have a volume that is ten times larger than in
the high-density regions. Without mass matching, the boundaries between
materials can become unstable and lead to unphysical instabilities.
In Kull, the zones are volume-matched, meaning that the mass of the
cells in the low-density region is ten times lower than the mass of
the cells in the high-density regions. These differences imply there
are regional resolution differences between the codes, though the
number of total elements (points vs. zones) is the same in each code.
The problem is run to a final time of 7.0 sh. 

\begin{figure}
\centering{}\subfloat[Radiation energy]{\includegraphics[width=0.96\textwidth]{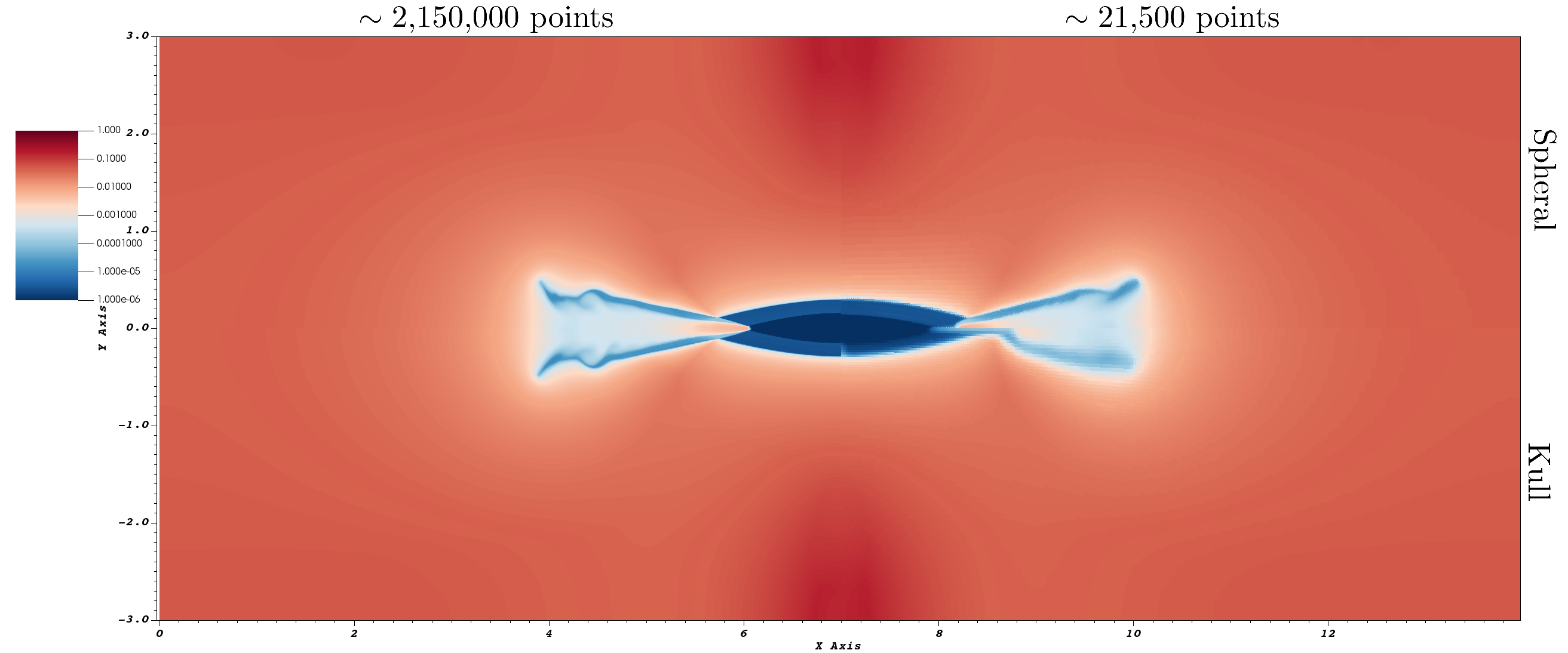}

}

\subfloat[Material energy]{\includegraphics[width=0.96\textwidth]{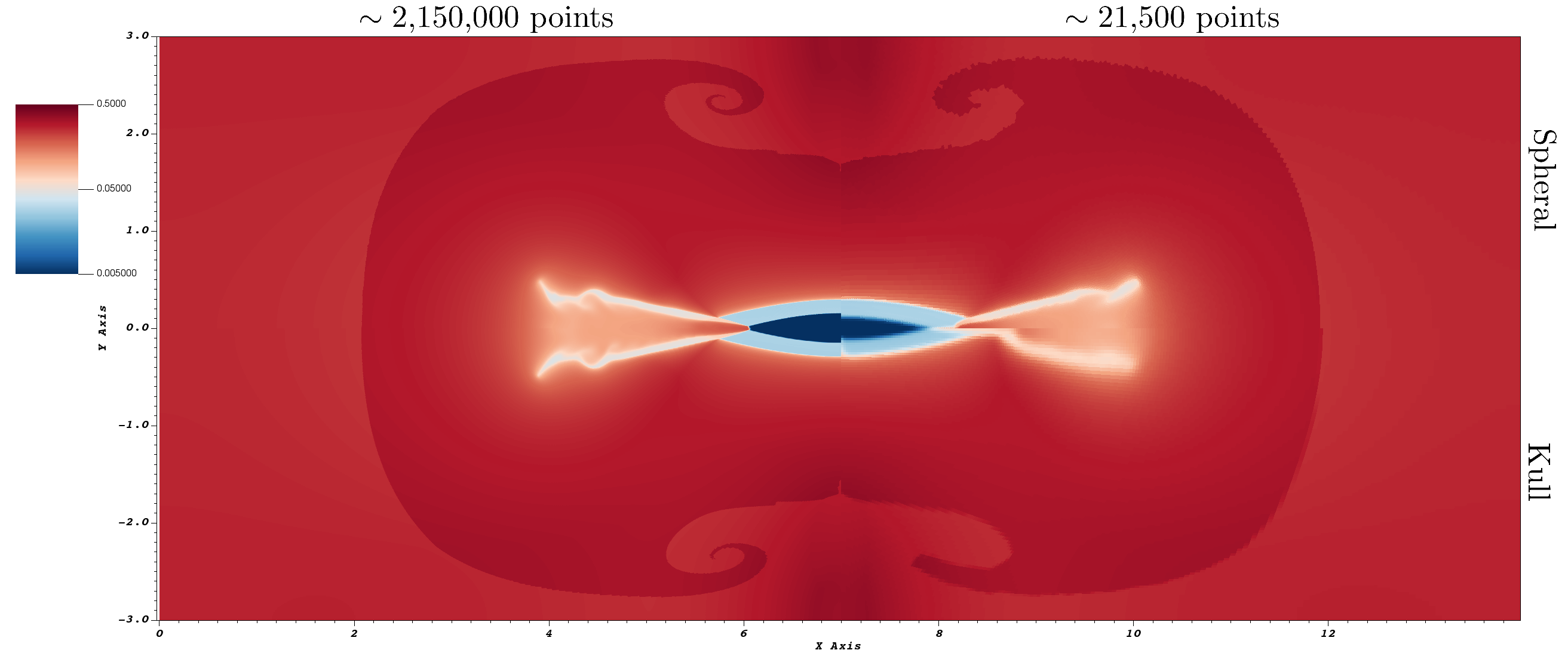}

}

\subfloat[Density]{\includegraphics[width=0.96\textwidth]{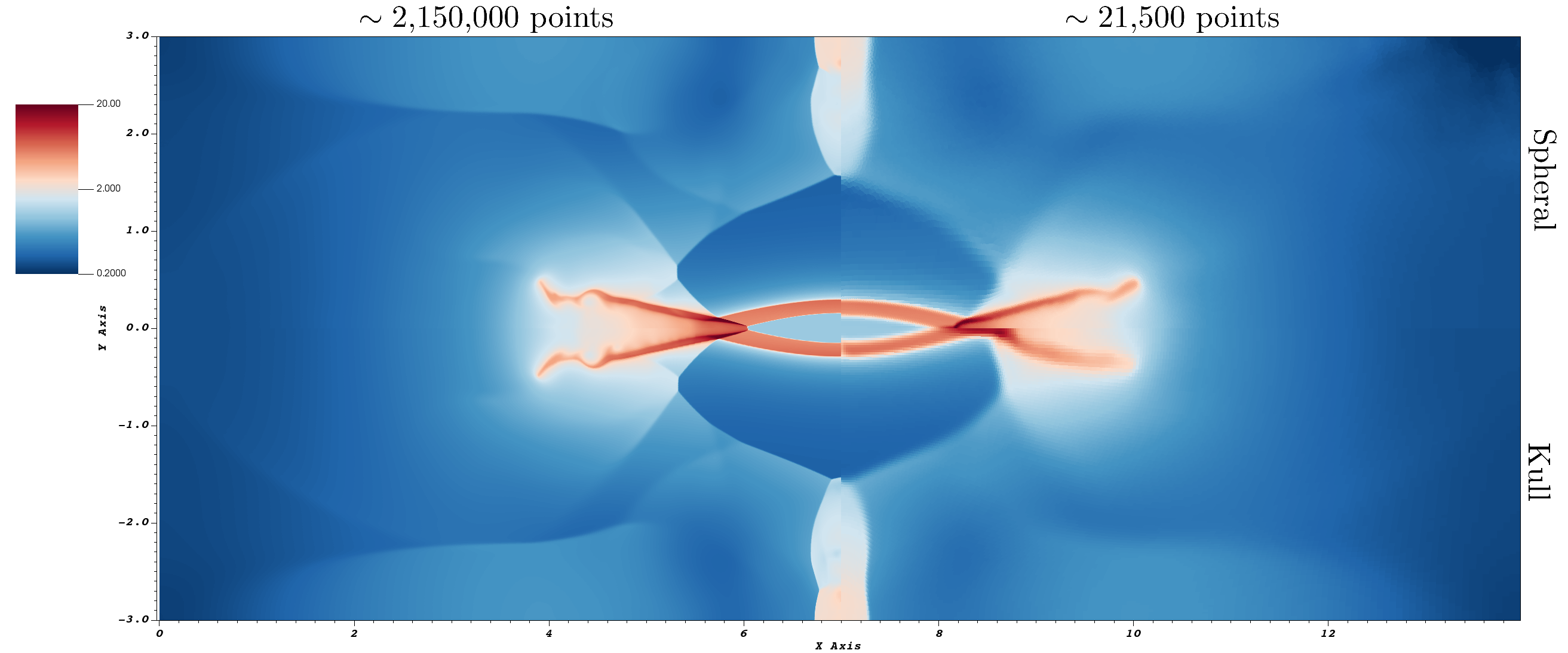}

}

\caption{Comparison of four test cases for the triple point problem at 7 sh.
For each plot, the top half is Spheral and the bottom half is Kull.
The left half is the high-resolution version, and the right half is
the low-resolution version. This figure is also available as a video. }
\label{fig:triple-solution}
\end{figure}

Figure \ref{fig:triple-solution} shows a comparison of the results
for Spheral and Kull. The first visible feature of the problem is
a radiation wave traveling in the $+\hat{x}$ direction through the
upper low-density region. This causes the surface of the lower region
to ablate upwards, sending an impulse in the $-\hat{y}$ direction.
Meanwhile, the high-pressure region creates a shock in the $+\hat{x}$
direction, which meets the shock formed due to the ablation. At the
end time, the most prominent features are the roll-up due to the Kelvin-Helmholtz
instability and the trailing tails of high-density material left from
the colliding shocks. For the high-resolution cases, the Spheral and
Kull solutions are nearly indistinguishable. For the low-resolution
cases, the arbitrary Lagrangian-Eulerian remap in Kull leads to a
very diffuse result, while the limited number of particles in Spheral
leads to shock boundaries that are smoothed compared to the high-resolution
cases. 

\begin{figure}
\centering{}\includegraphics[width=0.96\textwidth]{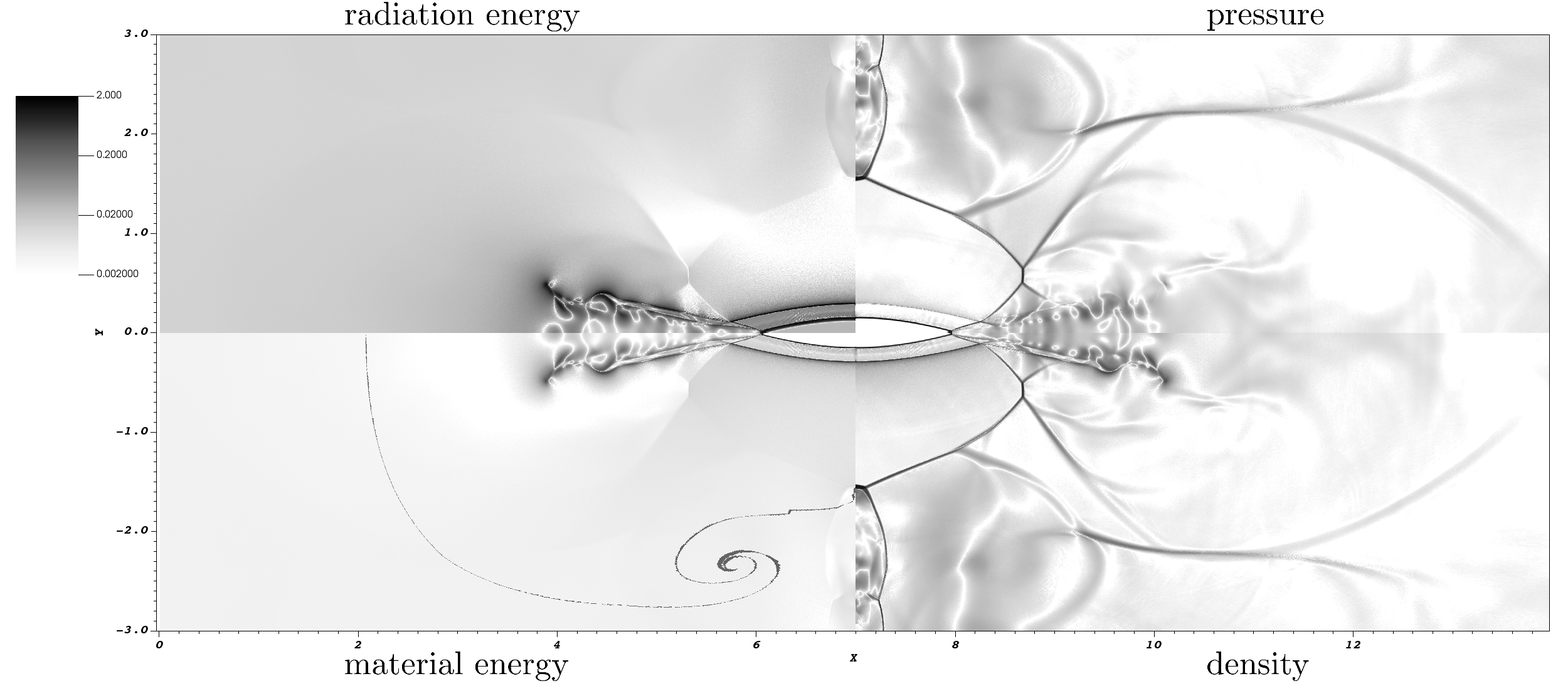}\caption{The difference between Spheral and Kull state variables, divided by
the average {[}Eq. (\ref{eq:kull-sph-diff}){]}, for the triple point
problem run with around 2,150,000 points or cells at 7 sh. This figure
is also available as a video. }
\label{fig:triple-difference}
\end{figure}

The difference between Kull and Spheral for the high-resolution cases,
defined as 
\begin{equation}
\text{diff}_{u}=\frac{\left|u_{\text{kull}}-u_{\text{spheral}}\right|}{\frac{1}{2}\left(u_{\text{kull}}+u_{\text{spheral}}\right)},\label{eq:kull-sph-diff}
\end{equation}
where $u$ is the radiation energy, material energy, pressure, or
density, is shown in Fig. \ref{fig:triple-difference}. The difference
is between a fraction of a percent and ten percent in most of the
domain for all the state variables shown. The most visible differences
are in the shock and wave timings, where the Kull radiation wave and
the Kull shocks are faster than those in Spheral. This leads to large
relative differences of up to 200 percent where the variables have
low values in pre-shock regions. The post-shock regions agree within
two percent in most cases. As the radiation energy density ranges
over six orders of magnitude, small differences in the timing lead
to differences that appear larger there than for the other state variables.
The interfaces between the three materials are very similar in the
two codes, and the main difference is in the material energy, where
the timing difference between the two is most visible. 

The relative difference between the initial and final energy, including
kinetic, internal, and radiation, is below 0.001 for both Kull and
Spheral, as shown in Table \ref{tab:triple-comparison}. The number
of steps for the low-resolution cases with around 21,500 points is
similar for the two codes at around 3,200. For the high-resolution
case with around 2,150,000 points, Spheral took 22,244 steps and Kull
took 152,843 steps, which is mostly due to deforming zones that decrease
the time step though the CFL condition. The wall time per time step
is about 2.5 times more in Spheral than Kull, mostly due to a higher
level of connectivity in the SPH diffusion matrix that slows the computation
of the multigrid preconditioner in Spheral. 

\subsection{Ablation problem\label{subsec:ablation}}

A target for inertial confinement fusion (ICF) \cite{nuckolls1972laser,lindl1995development}
generally consists of an outer spherical shell (or ablator) and an
inner shell made from deuterium-tritium (DT) ice, which contains a
low-density DT gas. A radiation source heats up the ablator, which
causes it to expand outward near the surface. Due to conservation
of momentum, this causes an implosion that compresses the DT ice and
gas. Under appropriate conditions, this leads to fusion of the DT.
For this problem, the physics of ICF are simplified to allow for comparison
of the SPH discretization in Spheral with the mesh-based discretization
in Kull. Unlike an actual ICF simulation, there is no radiation source
outside of the initial conditions and there is no thermonuclear burn. 

\begin{figure}
\begin{centering}
\includegraphics[width=0.6\textwidth]{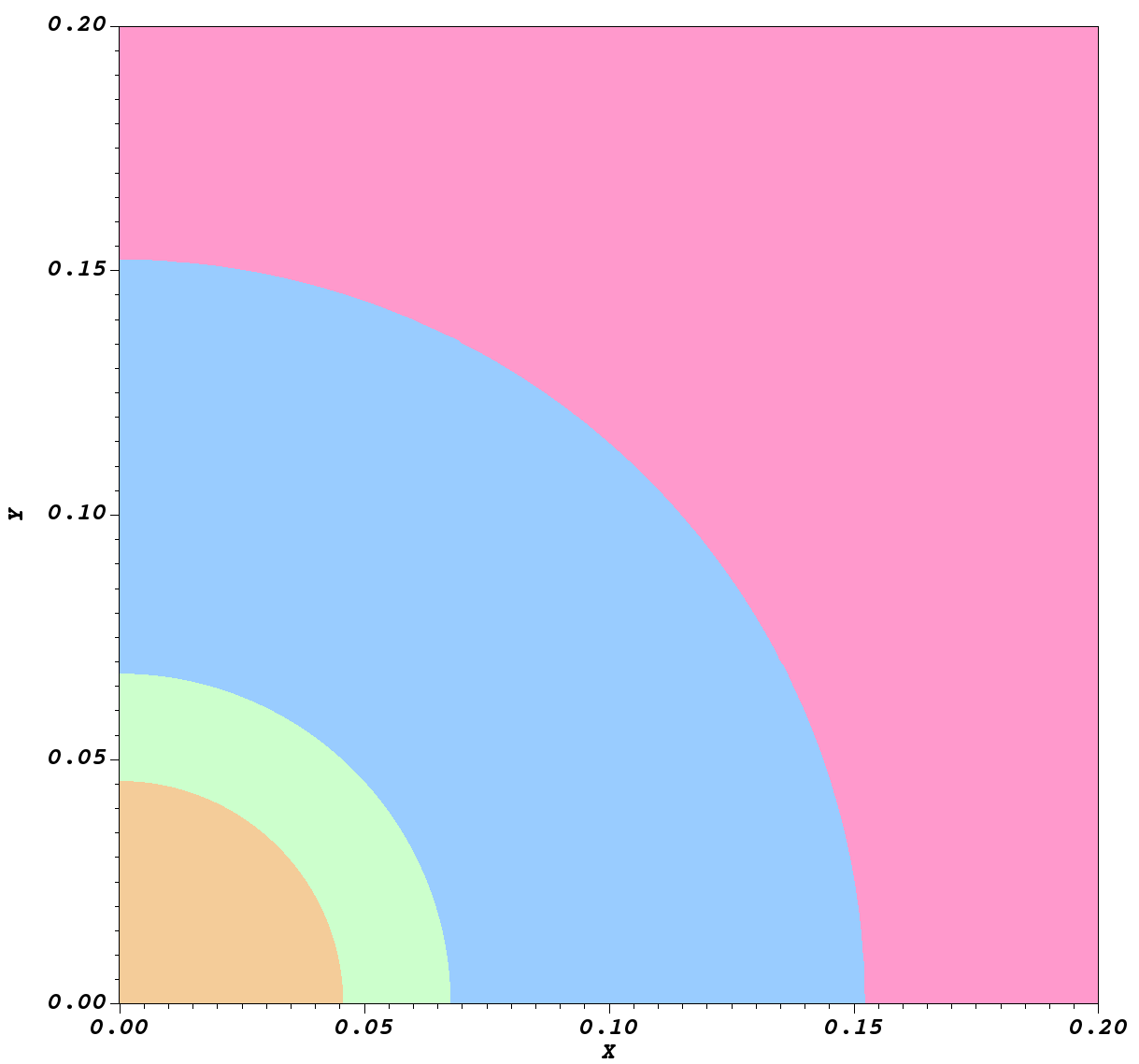}
\par\end{centering}
\caption{Geometry for the ablation problem. From the origin out, the regions
are DT gas, DT ice, CH 1\% Si plastic, and helium gas. In 2D, the
shells are infinite cylinders, while in 3D, the shells are spheres.}
\label{fig:ablation-geom}
\end{figure}

The capsule is modeled with a DT gas, an inner shell of DT ice, and
an outer plastic shell (CH 1\% Si) surrounded by helium gas \cite{tipton2016icf}.
In 2D, the shells are circles (or infinite cylinders) and the helium
is bounded by a square, while in 3D, the shells are spheres and the
helium is bounded by a cube. Figure \ref{fig:ablation-geom} shows
the initial geometry in 2D, or a slice through the geometry at $z=0$
in 3D. The problem is run with quadrant symmetry in 2D and octant
symmetry in 3D, with reflecting boundary surfaces along the coordinate
axes and at the edges of the square or cube. The unit system is identical
to the one in Sec. \ref{subsec:triple-point}. The initial parameters
are listed in Table \ref{tab:ablation-initial}. The opacities are
least squares fits of tabulated opacity data \cite{kumbera2002opacity}.
To make mass matching of the points easier, the initial densities
of the DT gas and the helium are around 100 times larger than what
would be expected in a standard ICF capsule. The material temperature
and the radiation temperature {[}Eq. (\ref{eq:rad-temperature}){]}
are set equal at the starting time. In 2D, the codes are run until
1.75 sh, while in 3D, the codes are run until 0.7 sh. 

The number of points used in each code is listed in Table \ref{tab:ablation-spatial}.
In Kull, the initial mesh is created with a square (2D) or cube (3D)
meshed region in the center of the DT gas sphere with rounded edges,
with a half-length equal to half of the radius of the DT gas region.
Zones are then placed radially outward from there, with the resolution
in angle fixed based on the number of regions in the square or cube.
The zones are mass matched across interfaces, which leads to zones
in the ice and ablator that are much smaller in the radial dimension
than in the swept angular dimensions. In Spheral, the number of radial
points for each region is chosen based on mass matching. The points
are initialized on concentric rings, with azimuthal spacing set equal
to the local radial spacing. The radial spacing is ratioed, with the
ratio defined as the radial distance at the outside radius over the
radial distance at the inside radius. The ratios are chosen in 2D
such that the points have smoothing lengths that do not differ drastically
at material boundaries. Spheral supports anisotropic point spacing
through adaptive SPH \cite{owen1998adaptive,Owen2010}, which would
decrease the number of points needed for a given radial resolution,
but to keep these examples consistent with standard SPH, this is not
employed here. 

\begin{figure}
\subfloat[Material energy]{\includegraphics[width=0.8\textwidth]{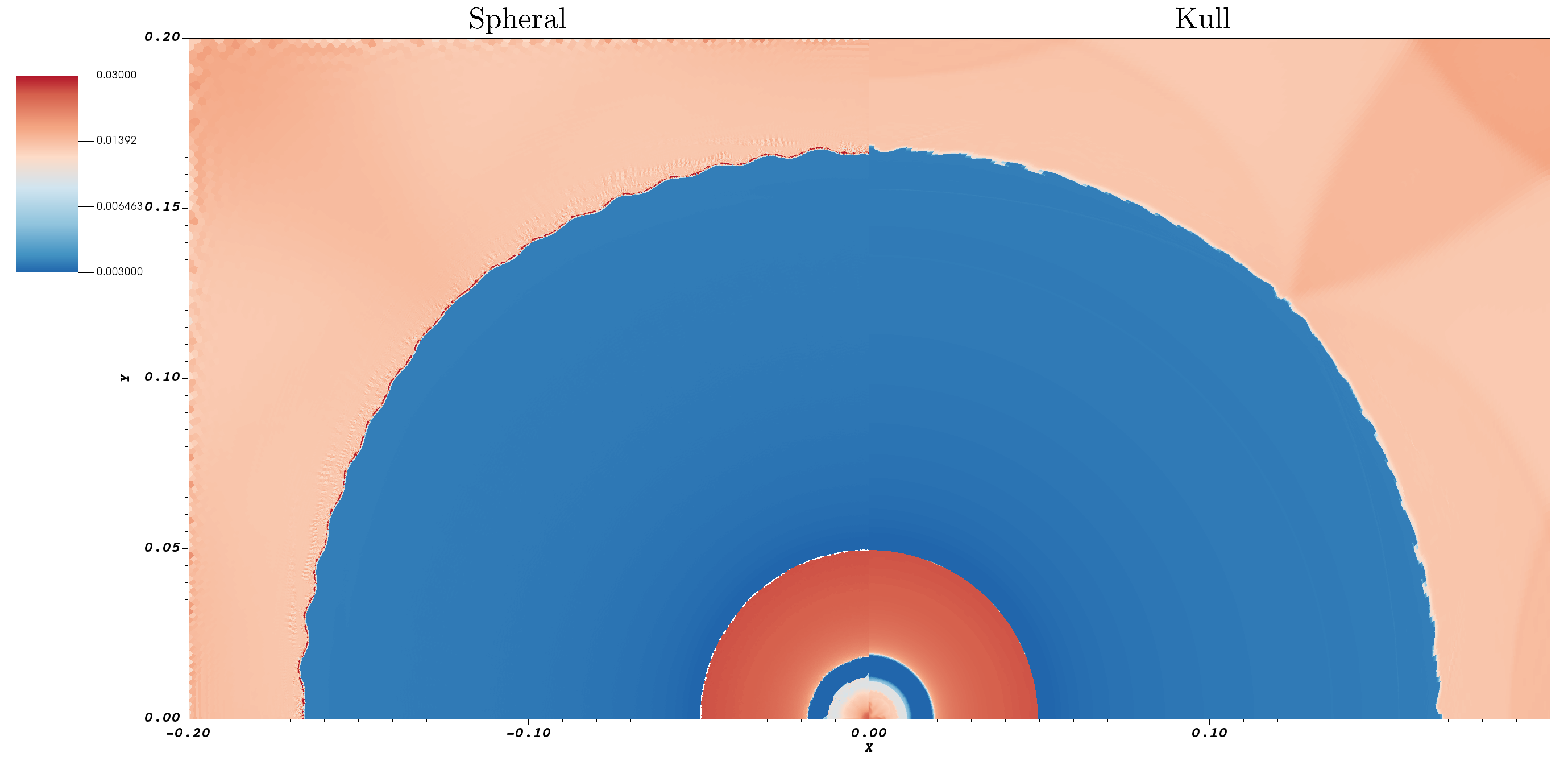}

}

\subfloat[Density]{\includegraphics[width=0.8\textwidth]{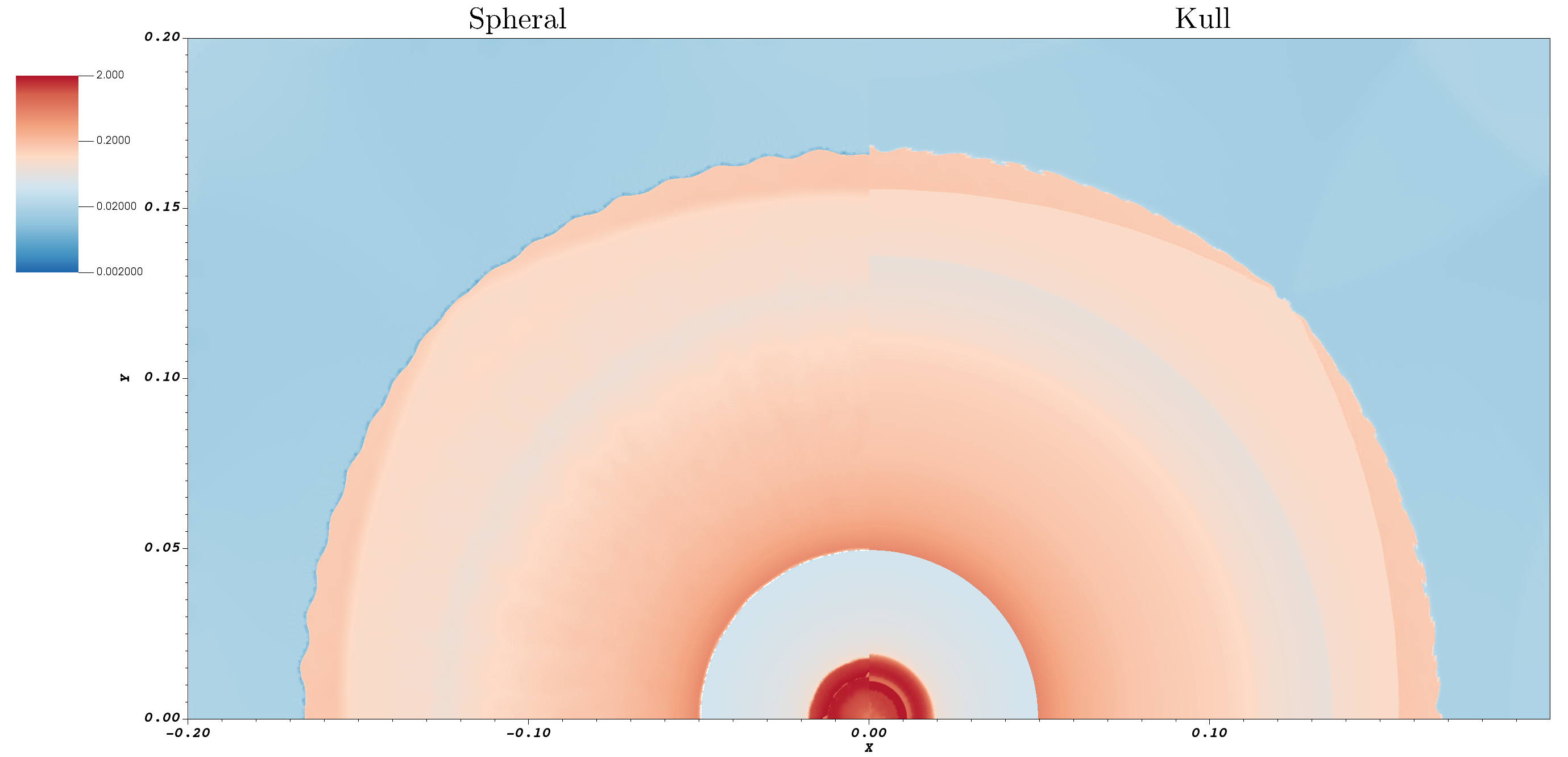}

}

\subfloat[Velocity]{\includegraphics[width=0.8\textwidth]{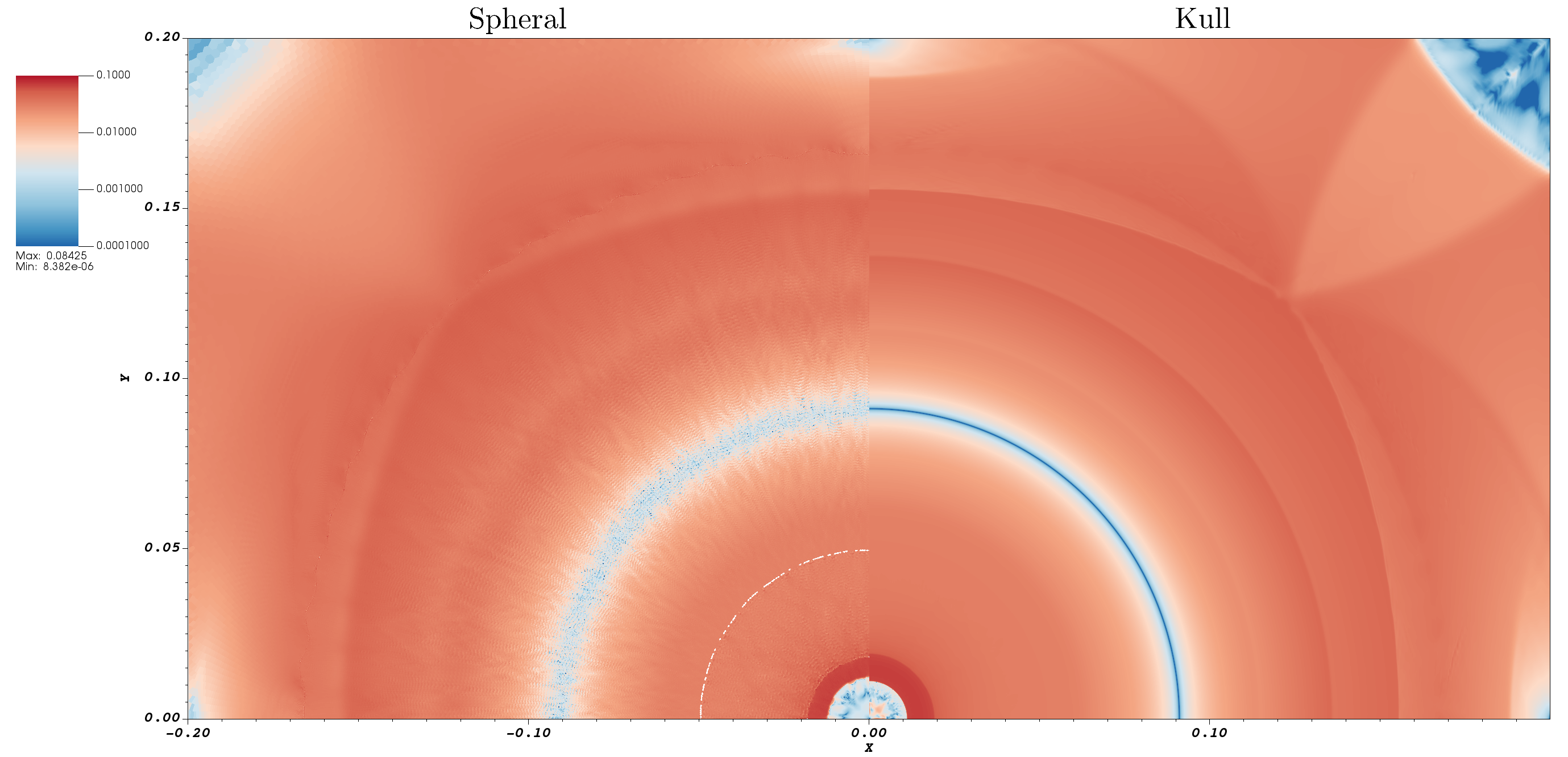}

}

\caption{Comparison of Kull and Spheral for the ablation problem in 2D at 1.2
sh, at maximum compression of the gas. See Table \ref{tab:ablation-spatial}
for the spatial discretization information. This figure is also available
as a video. }
\label{fig:ablation-2d-comparison}
\end{figure}

The comparison results for 2D are shown in Fig. \ref{fig:ablation-2d-comparison}.
At the initial time, the radiation in the helium gas heats the surface
of the ablator. As the ablator heats up, its opacities decrease, letting
the radiation penetrate further into the ablator. The rate at which
this occurs is governed by the thermal radiative transfer equations.
The ablation of the outer surface of the capsule creates an inward-moving
shock. This shock travels through the DT ice and eventually compresses
the DT gas. The shock travels through the DT gas and converges at
the center of the problem before the DT ice has stopped traveling
inward. The shock in the gas rebounds and bounces back off the ice
before the inward momentum of the ice is halted and the gas reaches
peak densities. The DT gas then pushes back against the DT ice, which
causes Rayleigh-Taylor instabilities at the gas-ice interface. 

The ablation rate of the capsule in the two codes is very similar.
The difference in the position of the ablator-ice interface, divided
by the initial radius of the interface, is around 0.15 percent (0.0001
cm) for most of the simulation, and is at most 1.5 percent (0.001
cm) near the end of the simulation. The ice-gas interface has similar
differences until the Rayleigh-Taylor instability develops, which
makes visual comparison difficult. The velocity plot shows the Spheral
shocks in the gas leading by about 0.02 sh, or 1.1 percent of the
total simulation time. The velocities between the codes are otherwise
very similar. The thermal energy and density are also very similar,
with the exception of the shape of the Rayleigh-Taylor instability
near the end time, which may be seeded by asymmetric initial conditions.
Table \ref{tab:ablation-timing} shows the timing and energy balance
information for each case. The computational cost for the two codes
is similar, as Spheral is more expensive per time step but takes fewer
time steps. The relative difference between the initial and final
energies is approximately the same between the two codes, at about
0.001 for Spheral and 0.002 for Kull. 

\begin{figure}
\subfloat[Material energy]{\includegraphics[width=0.8\textwidth]{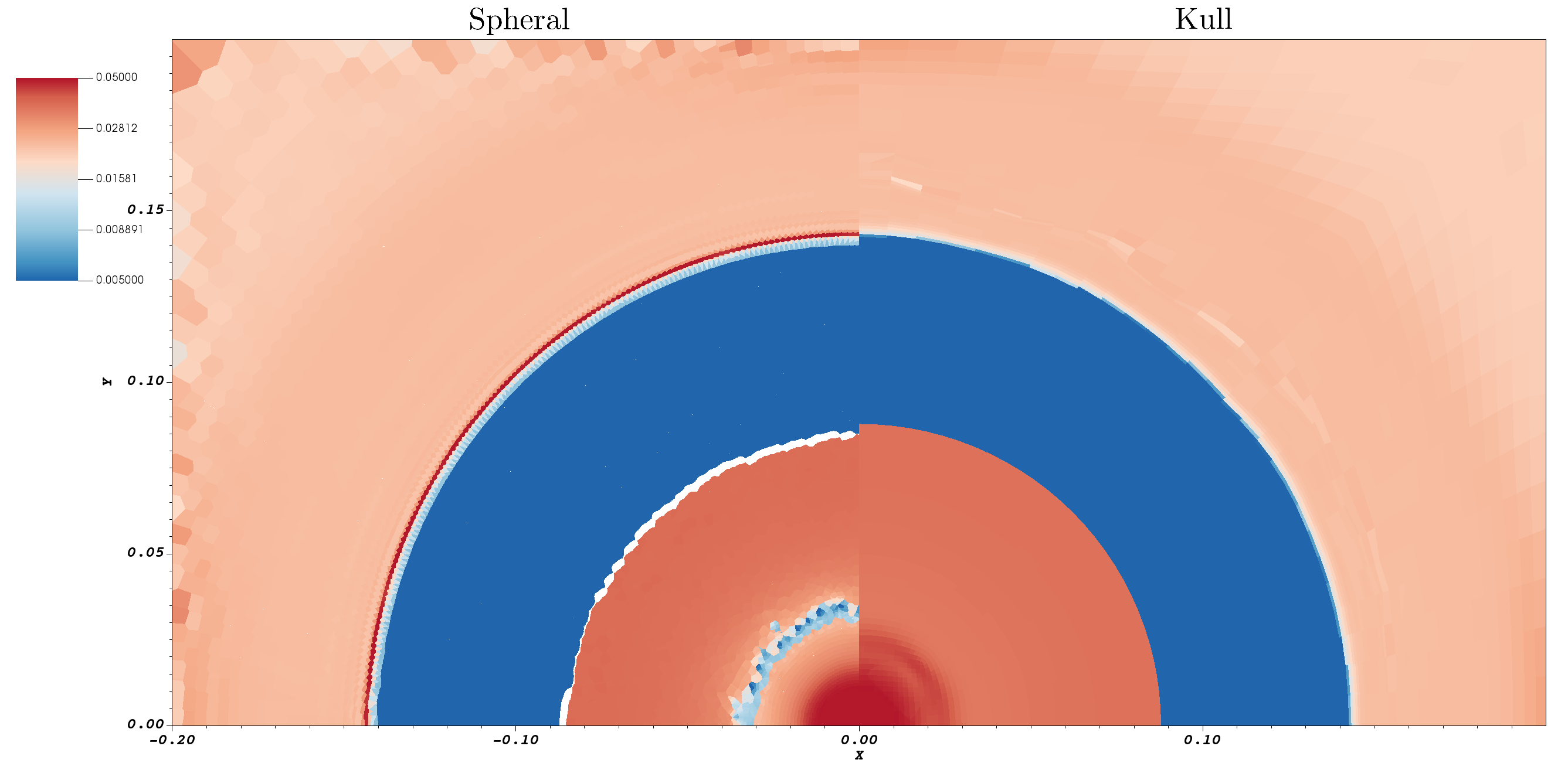}

}

\subfloat[Density]{\includegraphics[width=0.8\textwidth]{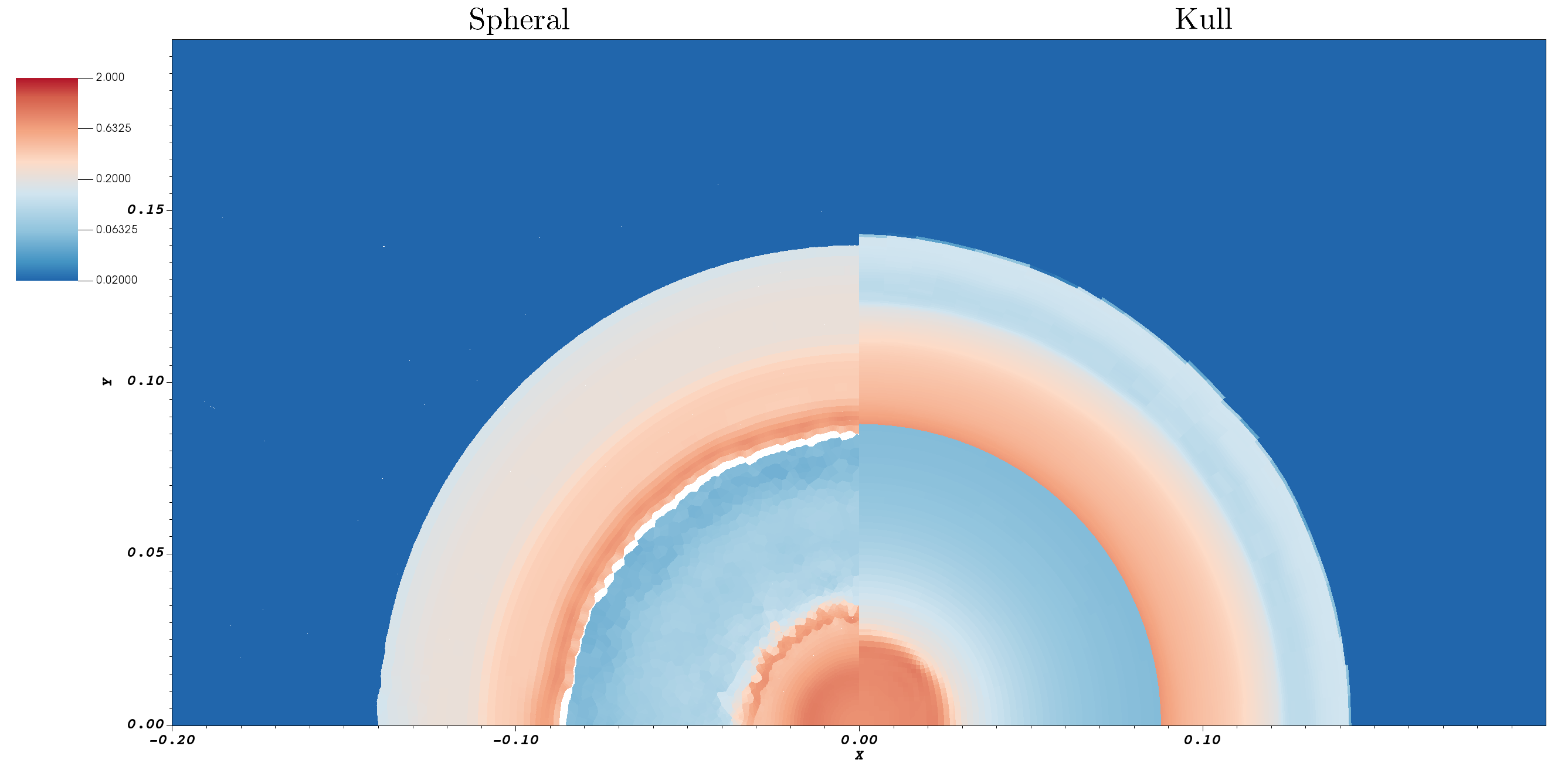}

}

\subfloat[Velocity]{\includegraphics[width=0.8\textwidth]{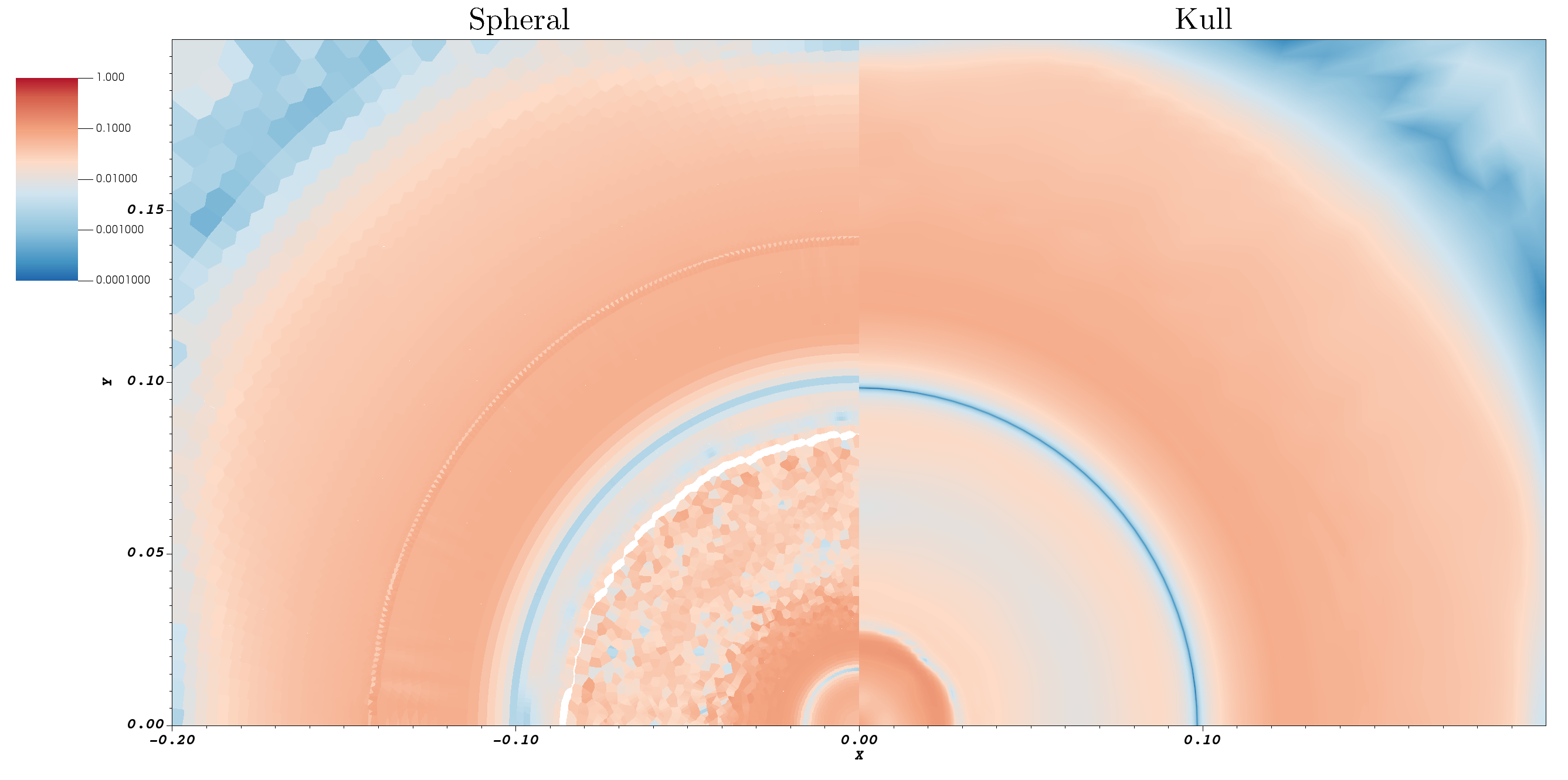}

}

\caption{Comparison of Kull and Spheral for the ablation problem in 3D at 0.48
sh. See Table \ref{tab:ablation-spatial} for the spatial discretization
information. This figure is also available as a video. }
\label{fig:ablation-3d-comparison}
\end{figure}

Figure \ref{fig:ablation-3d-comparison} shows the results in 3D,
which are not as well resolved as in 2D due to computational cost
(see Table \ref{tab:ablation-timing}), and because of this has only
a fraction of the radial zones of the 2D calculation (see Table \ref{tab:ablation-spatial}).
Despite having fewer points, the radial direction is more finely resolved
in Kull due to anisotropic cell lengths. The results are not converged
for either code, meaning that they may change significantly if the
spatial resolution is increased. Because of this, the results are
less similar than in 2D. Compared to 2D, the volume of the DT gas
and ice is much smaller compared to the ablator volume, which makes
the implosion progress more quickly.

The ablator-ice interface position is around 5 percent (0.003 cm)
different for the two codes throughout the simulation. The ice-gas
interface differs by more, at around 5.3 percent (0.008 cm) during
the implosion and 9.3 percent (0.014 cm) after the gas rebounds. The
stagnation points for the two simulations are similar in the outer
regions. The shocks in Spheral are behind those in Kull by around
0.01 sh, or 1.4 percent of the total simulation time, but otherwise
have similar behavior. Due to the comparatively low resolution of
the 3D simulations, the gas-ice instability on shock rebound is not
visible. 

\section{Conclusions and future work}

In this work a meshfree radiation hydrodynamics discretization based
on smoothed particle hydrodynamics has been demonstrated. A careful
splitting of the radiation hydrodynamics equations allows for the
hydrodynamics to be solved using an explicit time advance and the
radiation to be solved using an implicit time advance. The thermal
radiative transfer equations are solved using the discretization from
Paper I, while the hydrodynamics equations are solved using standard
SPH methodology. This allows for fully meshless radiation hydrodynamics
with arbitrary opacities and equations of state. By coupling with
the efficient radiation solve from Paper I, the meshfree method has
competitive performance when compared with a contemporary ALE radiation
hydrodynamics code designed to study inertial confinement fusion.

The Lowrie shock problem in 1D has a semianalytic solution that allows
verification of the radiation hydrodynamics equations. The numeric
solution converges to the semianalytic solution with second-order
accuracy for both the Mach 2 shock and the Mach 45 shock. Correct
implementation of the coupling terms between radiation and hydrodynamics
is essential for the Mach 45 problem, which has high radiation pressure
and tests both the propagation of radiation and hydrodynamic shocks. 

The triple point problem with radiation and an ablation problem based
off of an ICF capsule are tested against a mesh-based code, Kull.
In both cases, the main difference is minor timing differences in
the shock propagation. The triple point problem tests the ablation
rate of the top surface and the effects of radiation on the hydrodynamics.
The meshless results for a highly-converged case are visually indistinguishable
from the mesh-based code, and the difference between the two codes
apart from the shock timing differences is between a fraction of a
percent and 10 percent throughout the domain. For the ablation problem,
the interface positions agree to within 0.15 to 1.5 percent to a percent
in 2D and within 5.3 to 9.3 percent in 3D. For both problems, the
meshless code runs with similar speed to the mesh-based code due to
a smaller number of time steps required by the hydrodynamics in the
meshless code. 

As stated in Paper I, some improvements can be made to the diffusion
discretization to ensure zeroth-order consistency, either based on
the reproducing kernel particle method \cite{liu1995reproducing}
or moving least squares particle hydrodynamics \cite{dilts1999moving,dilts2000moving}.
For accuracy in energy, a multigroup treatment would be needed, preferably
coupled to an acceleration method such as linear multifrequency-grey
acceleration \cite{morel1985synthetic}. Finally, other angular treatments,
such as spherical harmonics or discrete ordinates transport, could
be considered for problems with small opacities or where radiation
is anisotropic. 

\section*{Acknowledgements}

The authors wish to acknowledge the assistance of Nick Gentile in
deriving coupling terms for the radiation hydrodynamics equations.

This work was performed under the auspices of the U.S. Department
of Energy by Lawrence Livermore National Laboratory under Contract
DE-AC52-07NA27344. This document was prepared as an account of work
sponsored by an agency of the United States government. Neither the
United States government nor Lawrence Livermore National Security,
LLC, nor any of their employees makes any warranty, expressed or implied,
or assumes any legal liability or responsibility for the accuracy,
completeness, or usefulness of any information, apparatus, product,
or process disclosed, or represents that its use would not infringe
privately owned rights. Reference herein to any specific commercial
product, process, or service by trade name, trademark, manufacturer,
or otherwise does not necessarily constitute or imply its endorsement,
recommendation, or favoring by the United States government or Lawrence
Livermore National Security, LLC. The views and opinions of authors
expressed herein do not necessarily state or reflect those of the
United States government or Lawrence Livermore National Security,
LLC, and shall not be used for advertising or product endorsement
purposes. LLNL-JRNL-799714. 

\bibliographystyle{unsrt}
\bibliography{sph_rad_hydro_refs}

\pagebreak{}

\begin{algorithm}
\begin{algorithmic}[1]
\State{$t=0$}
\State{set initial conditions}
\While{$t$ < end time}
\State{evaluate $\sigma_a,\sigma_s,\lambda$ using initial state}
\State{perform the explicit hydrodynamics update to get $\rho^n,v^n,e^n_{\text{hydro}}$}
\State{\qquad calculate radiation momentum using $E^{n-1}$}
\State{update the radiation energy density to $E^n$ using Eq. \eqref{eq:radiation-work}}
\State{\qquad store as $E^{n-1}$ for the radiation update}
\State{evaluate $\sigma_a,\sigma_s,\lambda,c_v,f$ using the values from the hydrodynamics update}
\State{calculate the net energy gain over the hydrodynamics update using Eq. \eqref{eq:material-energy-change}}
\State{perform the material/radiation update as described in Paper I, Alg. 1}
\State{\qquad use net energy gain as material energy source}
\State{$t = t + \Delta t^n$}
\EndWhile
\end{algorithmic}

\caption{Time integration procedure for smoothed particle radiation hydrodynamics.}
\label{alg:rad-hydro-time}
\end{algorithm}

\begin{table}
\begin{centering}
\begin{tabular}{|c|c|c|}
\hline 
 & Mach 2 & Mach 45\tabularnewline
\hline 
\hline 
$c_{v}$ ($\text{erg}\cdot\text{g}^{-1}\text{K}^{-1}$) & $1.911373*10^{8}$ & $1.246685\times10^{8}$\tabularnewline
\hline 
$\sigma_{a}$ ($\text{cm}^{-1}$) & $5.773503\times10^{2}$ & $7.565304\times10^{22}\rho^{2}T^{-3.5}$\tabularnewline
\hline 
$\sigma_{s}$ ($\text{cm}^{-1}$) & 0.0 & $0.4006\rho$\tabularnewline
\hline 
$\gamma$ (ratio of specific heats) & 5/3 & 5/3\tabularnewline
\hline 
$\mu$ (molecular weight) & 1.0 & 1.0\tabularnewline
\hline 
$\rho$ pre-shock ($\text{g}\cdot\text{cm}^{-3}$) & 1.0 & 1.0\tabularnewline
\hline 
$\rho$ post-shock ($\text{g}\cdot\text{cm}^{-3}$) & 2.286075 & 6.426142\tabularnewline
\hline 
$T$ pre-shock (K) & $1.410643\times10^{6}$ & $1.160452\times10^{6}$\tabularnewline
\hline 
$T$ post-shock (K) & $2.930711\times10^{6}$ & $8.69881\times10^{7}$\tabularnewline
\hline 
$v$ pre-shock ($\text{cm}\cdot\text{s}^{-1}$) & 0.0 & 0.0\tabularnewline
\hline 
$v$ post-shock ($\text{cm}\cdot\text{s}^{-1}$) & $-3.461705\times10^{7}$ & $-5.705363\times10^{8}$\tabularnewline
\hline 
End time (s) & $1.733250\times10^{-9}$ & $3.505474\times10^{-6}$\tabularnewline
\hline 
Distance traveled (cm) & $0.06$ & $2000.0$\tabularnewline
\hline 
Domain (cm) & $0.0\leq x\leq0.12$ & $0.0\leq x\leq2500.0$\tabularnewline
\hline 
\end{tabular}
\par\end{centering}
\caption{Description of the Lowrie shock problem for Mach 2 and Mach 45. For
opacity evaluation, the density is in $\text{g}\cdot\text{cm}^{-3}$,
while the temperature is in K.}
\label{tab:lowrie-parameters}
\end{table}

\begin{table}
\begin{centering}
\begin{tabular}{|c|c|c|c|c|c|c|c|}
\hline 
 & Points/Cells & Procs & Points/proc & Steps & Wall time & Time per step & Energy error\tabularnewline
\hline 
\hline 
\multirow{2}{*}{Spheral} & 21,410 & 36 & 595 & 3253 & 1,641 & 0.505 & $-7.466\times10^{-4}$\tabularnewline
\cline{2-8} \cline{3-8} \cline{4-8} \cline{5-8} \cline{6-8} \cline{7-8} \cline{8-8} 
 & 2,149,550 & 1152 & 1,866 & 22,244 & 62,259 & 2.799 & $-1.137\times10^{-4}$\tabularnewline
\hline 
\multirow{2}{*}{Kull} & 21,504 & 36 & 598 & 3101 & 563 & 0.182 & $-7.449\times10^{-4}$\tabularnewline
\cline{2-8} \cline{3-8} \cline{4-8} \cline{5-8} \cline{6-8} \cline{7-8} \cline{8-8} 
 & 2,150,400 & 1152 & 1,867 & 152,843 & 170,800 & 1.117 & $2.993\times10^{-4}$\tabularnewline
\hline 
\end{tabular}
\par\end{centering}
\caption{Run parameters and comparison of computational cost for the triple
point problem. The energy error is the initial minus the final energy,
divided by the initial energy.}
\label{tab:triple-comparison}
\end{table}

\begin{table}
\begin{centering}
\begin{tabular}{|c|c|c|c|c|c|c|c|}
\hline 
Material & Outer radius (cm) & Initial $\rho$ ($\text{g}\cdot\text{cm}^{-3}$) & Initial $T$ (keV) & $\gamma$ & $\mu$ & $\sigma_{a}$ (cm$^{-1}$) & $\sigma_{s}$ (cm$^{-1}$)\tabularnewline
\hline 
\hline 
DT Gas & 0.08628 & $3.0\times10^{-2}$ & $1.7\times10^{-6}$ & 1.45 & 1.0 & \multirow{2}{*}{$83.0\rho^{1.56}T^{-2.5}$} & \multirow{4}{*}{1.0$\rho$}\tabularnewline
\cline{1-6} \cline{2-6} \cline{3-6} \cline{4-6} \cline{5-6} \cline{6-6} 
DT Ice & 0.0943 & $0.25$ & $4.2\times10^{-7}$ & 1.45 & 1.0 &  & \tabularnewline
\cline{1-7} \cline{2-7} \cline{3-7} \cline{4-7} \cline{5-7} \cline{6-7} \cline{7-7} 
CH 1\% Si & 0.1108 & $1.073$ & $8.6\times10^{-6}$ & 1.3 & 13.0 & $5.7\rho^{1.62}T^{-2.7}$ & \tabularnewline
\cline{1-7} \cline{2-7} \cline{3-7} \cline{4-7} \cline{5-7} \cline{6-7} \cline{7-7} 
Helium & 0.2 & $1.0\times10^{-2}$ & $0.4$ & 1.66 & 2.0 & $0.30\rho^{1.73}T^{-3.3}$ & \tabularnewline
\hline 
\end{tabular}
\par\end{centering}
\caption{Input parameters for ablation problem. For opacity evaluation, density
is in $\text{g}\cdot\text{cm}^{-3}$ and temperature is in keV. The
listed radius for helium is the half-length of the square or cube.
The $(\gamma,\mu)$ columns are the (ratio of specific heats, mean
molecular weight) for the gamma-law gas equations of state in each
region.}
\label{tab:ablation-initial}
\end{table}

\begin{table}
\begin{centering}
\begin{tabular}{|c|c|c|c|c|c|c|}
\hline 
Code & Dimension & DT Gas & DT Ice & CH 1\% Si & Helium & Points/cells\tabularnewline
\hline 
\hline 
\multirow{2}{*}{Spheral} & \multirow{1}{*}{2} & 134 (0.5) & 36 (0.5) & 154 (1.0) & 155 (20.0) & 328,465\tabularnewline
\cline{2-7} \cline{3-7} \cline{4-7} \cline{5-7} \cline{6-7} \cline{7-7} 
 & \multirow{1}{*}{3} & 38 (1.0) & 7 (1.0) & 23 (1.0) & 52 (10.0) & 1,058,392\tabularnewline
\hline 
\multirow{2}{*}{Kull} & \multirow{1}{*}{2} & 96, 96 & 144 & 1296 & 128 & 328,704\tabularnewline
\cline{2-7} \cline{3-7} \cline{4-7} \cline{5-7} \cline{6-7} \cline{7-7} 
 & \multirow{1}{*}{3} & 14, 14 & 22 & 192 & 19 & 147,980\tabularnewline
\hline 
\end{tabular}
\par\end{centering}
\caption{Number of radial points or zones in each region for the ablation problem.
The numbers in parentheses for Spheral are the ratio of the radial
distance between points at the inner radius to the outer radius. In
the DT gas, Kull uses two parameters, the number of zones across a
square or cube at the center of the region and the number of radial
regions outside of it.}
\label{tab:ablation-spatial}
\end{table}

\begin{table}
\begin{centering}
\begin{tabular}{|c|c|c|c|c|c|c|c|}
\hline 
Code & Dimension & End time & Points/cells & Procs & Steps & Time per step & Energy error\tabularnewline
\hline 
\hline 
\multirow{2}{*}{Spheral} & \multirow{1}{*}{2} & 1.75 & 328,465 & 288 & 19,264 & 1.814 & $9.286\times10^{-4}$\tabularnewline
\cline{2-8} \cline{3-8} \cline{4-8} \cline{5-8} \cline{6-8} \cline{7-8} \cline{8-8} 
 & \multirow{1}{*}{3} & 0.7 & 1,058,392 & 1152 & 3,156 & 25.360 & Not recorded\tabularnewline
\hline 
\multirow{2}{*}{Kull} & \multirow{1}{*}{2} & 1.75 & 328,704 & 144 & 63,488 & 0.880 & $-2.416\times10^{-3}$\tabularnewline
\cline{2-8} \cline{3-8} \cline{4-8} \cline{5-8} \cline{6-8} \cline{7-8} \cline{8-8} 
 & \multirow{1}{*}{3} & 0.7 & 147,980 & 144 & 4,795 & 1.754 & $1.181\times10^{-3}$\tabularnewline
\hline 
\end{tabular}
\par\end{centering}
\caption{Timing and step information for the ablation problem. The energy error
is the initial minus the final energy, divided by the initial energy.}
\label{tab:ablation-timing}
\end{table}

\begin{figure}
\includegraphics[width=0.96\textwidth]{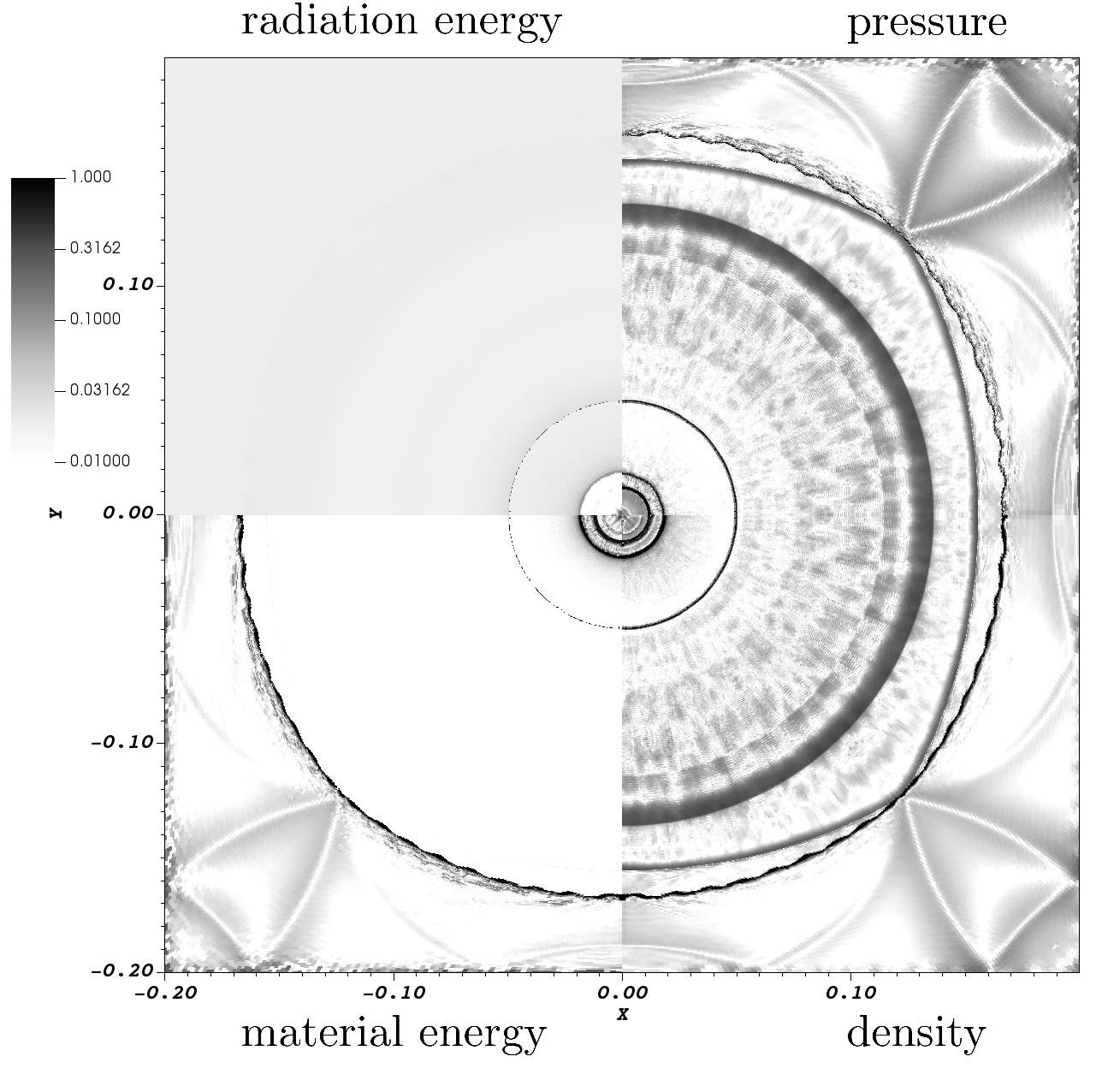}

\caption{The difference between Spheral and Kull state variables, divided by
the average {[}Eq. (\ref{eq:kull-sph-diff}){]}, for the ablation
problem in 2D at 1.2 sh. This figure is also available as a video. }
\label{fig:ablation-2d-difference}
\end{figure}

\end{document}